\documentclass{aa}
\usepackage{graphics,txfonts,amsmath}

\begin{document} 

\newcommand{\Hi}{\textup{H\,{\mdseries\textsc{i}}}}
\newcommand{\HI}{\textup{H\,{\mdseries\textsc{i}}}~}
\newcommand{\HII}{\textup{H\,{\mdseries\textsc{ii}}}}
\newcommand{\Kkms}{{K~km~s$^{-1}$}}
\newcommand{\kms}{{km~s$^{-1}$}}
\newcommand{\Htwo}{{H$_{2}$}\ }
\newcommand{\HaN}{{H$\alpha$+[N II]}\ }
\newcommand{\Ha}{{H$\alpha$}\ }
\newcommand{\coa}{{$^{12}$CO(1-0)}}
\newcommand{\cob}{{$^{12}$CO(2-1)}}
\newcommand{\Msun}{{M$_{\odot}$}}
\newcommand{\Lsun}{{L$_{\odot}$}}
\newcommand{\MHtwo}{{M$_{\rm H_2}$}}
\newcommand{\MHI}{{M$_{\rm H_I}$}}
\newcommand{\column}{{$10^{20}$~cm$^{-2}$~(K~km~s$^{-1}$)$^{-1}$}}
\newcommand{\vrel}{{v$_{\rm rel}$}}
\newcommand{\Rc}{{R$_{\rm c}$}}
\newcommand{\dv}{{$\Delta \rm v$}}
\newcommand{\Mvir}{{M$_{ \rm vir}$}}
\newcommand{\gmr}{($g$-$r$)}
\def\kms{{km~s$^{-1}$}}
\def\fmag{\hbox{$.\!\!^{\rm m}$}}
\def\degr{\hbox{$^\circ$}}
\def\arcmin{\hbox{$^\prime$}}
\def\arcsec{\hbox{$^{\prime\prime}$}}
\def\fdegr{\hbox{$.\!\!^\circ$}}
\def\farcmin{\hbox{$.\mkern-4mu^\prime$}}
\def\farcsec{\hbox{$.\!\!^{\prime\prime}$}}


 \title{Dorado and its member galaxies}

\subtitle{H$\alpha$ imaging of the group backbone}

   \author{R. Rampazzo
          \inst{1,2}
          \and
          {S. Ciroi}\inst{3}
                      \and
                P. Mazzei\inst{2}
              \and
      {F. Di Mille}\inst{4}
      \and
      {E. Congiu}\inst{4,5} 
      \and
            {A. Cattapan}\inst{3} 
      \and \\
       {L. Bianchi}\inst{6}
       \and 
    {E. Iodice}\inst{7}
       \and
    {A. Marino}\inst{2}               
\and 
{H. Plana}\inst{8}
              \and 
{J. Postma}\inst{9}
    \and
   {M. Spavone}\inst{7}
      }

   \institute{INAF-Osservatorio Astrofisico di Asiago,Via dell'Osservatorio 8, 36012 Asiago, Italy
              \email{roberto.rampazzo@inaf.it}
         \and
   {INAF-Osservatorio Astronomico di Padova}, {Vicolo dell'Osservatorio 5},  {35122 Padova}, {Italy}
        \and
        {Department of Physics and Astronomy  ``G. Galilei'', University of Padova}, {Vicolo dell'Osservatorio 3}, {35122 Padova}, {Italy}
                          \and
             {Las Campanas Observatory}, {Carnegie Institution of Washington}, {Colina El Pino Casilla 601, La Serena}, {Chile}
             \and
Departamento de Astronom\'{i}a, Universidad de Chile, Camino del Observatorio 1515, Las Condes, Santiago, Chile
\and
             {Dept. of Physics \& Astronomy}, {The Johns Hopkins University}, { 3400 N. Charles St., Baltimore}, {MD 21218, USA}
                         \and
             {INAF-Osservatorio Astronomico di Capodimonte}, {Salita Moiariello 16}, {80131 Napoli}, {Italy}          
       \and
        {Laborat\'orio de Astrof\'isica Te\'orica e Observacional}, {Universidade Estadual de Santa Cruz}, {45650-000 Ilh\'eus}, {BA}, {Brasil}  
       \and     
             {University of Calgary}, {2500 University Drive NW, Calgary}, {Alberta, Canada}
    }

   \date{Received ; accepted }

 
  \abstract
{Dorado is a nearby, rich and clumpy galaxy group that extends for several degrees 
 in the Southern Hemisphere. Although several studies have been dedicated to
define its members, their kinematics, hot and cold gas content, in particular \HI,  
their present star formation activity is yet unknown.}
{For the first time, we map the \Ha\ distribution as a possible indicator of star
formation activity of Dorado members a large fraction 
of which show interaction and merging signatures, regardless of their morphological type.}
{With the 2.5m du Pont and the 1m Swope telescopes we obtained  narrow-band,
 calibrated images of 14 galaxies, forming the backbone of
the group, mapping  \HaN\  down to few 10$^{-17}$ erg~cm$^{-2}$~s$^{-1}$~arcsec$^{-2}$.
We estimated the galaxy star formation rate from the \Ha\ fluxes, corrected 
for Galaxy foreground extinction and [N II] contamination.}   
{ \HaN\ emission has been detected in all galaxies.   
\HII\ regions clearly emerge in late-type galaxies, while in early-type galaxies the \HaN\ 
 emission is dominated by [N II], especially  in the central regions. 
However, \HII\ complexes are revealed in four early-type galaxies. 
Even in the compact group SGC 0414-5559, in the projected centre of Dorado,  
\HII\ regions are found both throughout the late-type galaxies
and in the very outskirts of early-type members.
Considering the Dorado group as a whole, we notice that
the \HaN\ equivalent width, a measure of the specific star formation,  increases with 
the morphological type, from early to late-type members, although it remains
lower that what observed in similar surveys of spiral galaxies.
The star formation rate of the spiral members is in the range of what observed in
similar galaxies surveys \citep{James2004}. However, in three spiral NGC 1536, PGC 75125
and IC 2058 the star formation rate is well below the median for their morphological classes. 
The star formation rate of some early-type members tends, at odds, to be higher than the average
derived from \HaN\ surveys of this morphological family.}
{We detected in \HaN\ all the early type galaxies observed and half of them show \HII\ regions in well shaped rings as well as 
in their outskirts. These findings suggest that ETGs in this group are not dead galaxies: 
their star formation has not shut down yet. Mechanisms such as gas stripping and gas accretion, 
through galaxy-galaxy interaction, seem relevant in modifying star formation in this 
evolutionary phase of Dorado.}


\titlerunning{H$\alpha$ imaging of the Dorado group backbone}
\authorrunning{Rampazzo et al.}
   \maketitle
%

\section{Introduction}
\label{Introduction}

Driven by gravitation, groups and their member galaxies co-evolve. 
During the group evolution, galaxies deeply transform their 
properties, as we start to learn from color-magnitude diagrams 
(CMDs hereafter) which provide a color snapshot of this process
\citep{Balogh2004,Baldry2004,Schawinski2007,Kaviraj2007}.
In CMDs, blue vs red galaxy populations trace 
the transition from an active to a more evolved or 
even passive phase \citep{Marino2010,Marino2013,Marino2016,Rampazzo2018}. 
Member galaxies modify both their star formation rate and morphology,
from star forming late-type (LTGs hereafter) to early-type (ETGs=Es+S0s hereafter), 
quenching their star formation \citep{Mazzei2014b}. 
Several physical mechanisms are believed to play a role in this galaxy 
transformation, according to the richness of their 
environments \citep[see e.g][]{Boselli2006,Boselli2014}.
Since velocity dispersions of galaxies in groups are comparable
to the velocity dispersion of stars in individual galaxies, both
interactions and merging are more favored as mechanisms than in clusters
\citep{Mamon1992}. Mergers can transform LTGs  into 
ETGs, that is both S0s and Ellipticals
\citep[see e.g.][]{Toomre1972,Barnes2002,Mazzei2014a,Mazzei2019} 
and quench star formation (SF hereafter) by ejecting the 
interstellar medium via starburst, AGN or shock-driven winds 
\citep[see e.g.][and references therein]{DiMatteo2015}.

\citet{Tal2009} presented a deep  imaging study, reaching 
$\mu_V$=27.7 mag arcsec$^{-2}$,  of a complete sample of luminous ETGs
(M$_B <$ -20) at distances 15-50 Mpc, selected from the
\citet{Tully1988a} catalog of nearby galaxies. They find that 73\% of
them show tidal disturbance signatures in their stellar bodies.
Concerning the relation between gravitational interaction signatures and
the galaxy environment they find that galaxies in clusters are less
perturbed than group and field galaxies. \citet{Tal2009}  concluded that
ETGs in groups and low-density environments continue to grow at the
present day through mostly dry mergers involving little star
formation. More recently, \citet{Rampazzo2020} found that $\approx$60\% of
isolated ETGs show shells structures, an unambiguous merging signature, 
extending  the view of intense galaxy transformation to very low 
density environments.

\medskip
In LTGs the \Ha\ flux is directly connected to the star 
formation but this is not the case for ETGs.
When \Ha\ started to be widely detected in ETGs central regions, using 
both spectroscopy \citep[][]{Heckman1980,Phillips1986} and imaging
\citep{Goudfrooij1994a,Goudfrooij1994b}, several mechanisms have been 
suggested to be at the origin of the emission in these galaxies considered
red-and-dead. The \Ha\ extended emission was thought 
to be connected with the hot gas phase giving rise to the strong X-ray emission
in ETGs \citep[see e.g.][and references therein]{Trinchieri1997}.
The nuclei of most ETGs show  low ionization nuclear emission
regions (LINER hereafter) properties that could explain the \Ha\ emission in the 
central regions. The role of Post Asymptotic Giant Branch (PAGB) stars has 
been also emphasized when the emission is more extended 
\citep[see e.g.][]{Annibali2010}. Both {\it GALEX}  
\citep{Jeong2009,Marino2011b} and {\it Spitzer} \citep[see e.g.][and references
therein]{Panuzzo2007,Vega2010,Panuzzo2011,Rampazzo2013}
observations have increasingly demonstrated that SF can also be found
in ETGs and can be at the origin of the \Ha\ emission both in the centre
and in the outer regions, often characterized by resonance rings. ETGs have been often
and often revealed also in \Hi, the fuel of the SF \citep[see e.g.][and references]{Serra2012}.
Recently, a \Ha\ survey of 147 ETGs, out of 260 in the {\it ATLAS$^{3D}$} 
sample has been performed by \citet{Gavazzi2018}.   Ninety-two ETGs were undetected. 
They are gas-free systems which lack a disk and exhibit passive spectra, even in their nuclei. 
Most (76\%) of their remaining ETGs are strong \Ha\ emitters and were found 
to be associated with low-mass  (M*$\approx$10$^{10}$ M$_\odot$) S0 galaxies, 
showing conspicuous gas (\HI\ + H$_2$) content, 
extended stellar disks and SF also in their nuclei.

\medskip

In  the above context, the study of the Dorado group, offers the possibility 
to investigate the gas rich perspective in studying accretion and merging events.
 Dorado is a \HI\ rich group, with a total \Hi\ mass of the group is 
	3.5$\times$10$^{10}$ M$_\odot$, nearly half of which concentrated in the 
	spiral member NGC 1566 \citep{Kilborn2005,Kilborn2009}. 
	These studies detected 13 galaxies including ETGs (NGC 1533, 
	NGC 1543, NGC 1596).  In addition, the \HI\ distribution offers indication of the strong
	interaction between Dorado members, that may re-fuel gas poor systems as ETGs.
	\citet{Ryan2003} revealed a vast system of \Hi\ around NGC 1533
	with tail connecting the galaxy to the IC2038/IC 2039 
	pair, located several kpcs away.
Indeed, irrespective of their morphological classification, Dorado members
show a significant fraction of galaxies,  from the centre to group
periphery, showing  either galaxy-galaxy interaction or merging
signatures.

The UV-optical Dorado CMD \citep[see Figure 1 in][]{Cattapan2019} is composed of 
both a red sequence, including several ETGs, and a 
Green Valley populated by numerous intermediate luminosity galaxy members.
Only NGC 1566, a bright grand design spiral, is still located in the Blue Cloud. 
Therefore, with respect to the rich  NGC 5486 group \citep[][]{Marino2016}, Dorado is in an
earlier and active evolutionary phase (see \S~\ref{Dorado-description}) 
and offers the possibility of investigating the evolutionary mechanisms in action.
Our study with {\it GALEX} of 
the early-type members, NGC 1533 and NGC 1543 \citep{Marino2011a,Rampazzo2017},
provide direct evidence of the presence of SF.
\citet{Mazzei2014a,Mazzei2014b} smoothed particle hydrodynamic 
(SPH) simulations with chemo-photometric implementation succeeded in reproducing 
the morphological, kinematical and the spectral energy distribution 
of two members of the Dorado group, NGC 1533 and NGC 1543, 
as a merger by-product \citep{Mazzei2019}.

\begin{figure*}
	\center
	{\includegraphics[width=14.5cm]{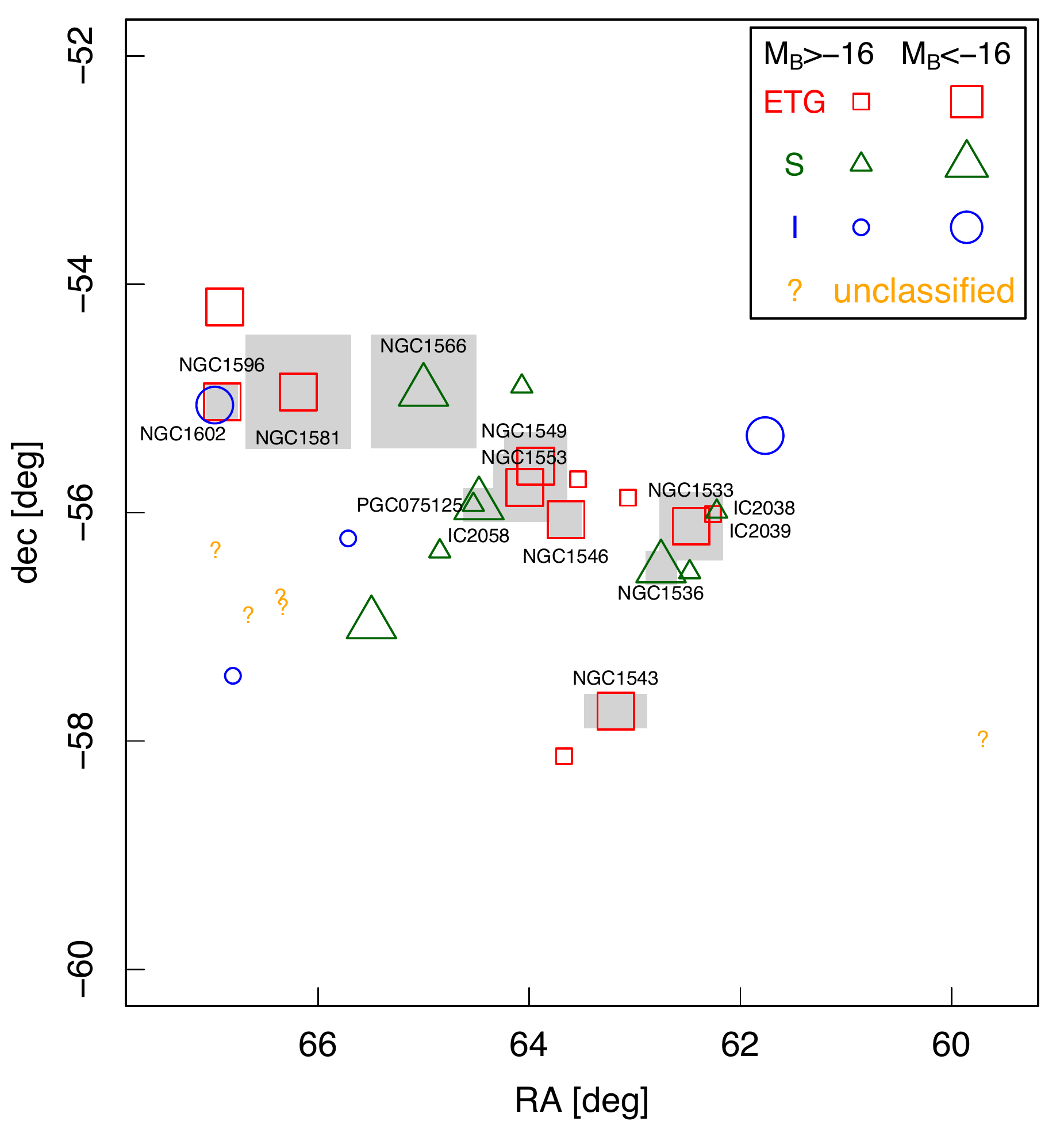}} \caption{
		Projected distribution of the Dorado group member
		galaxies (see Appendix~\ref{Dorado members}). 
		Galaxies labeled are those investigated in the present
		\HaN study. Galaxies are indicated according to their B-band 
		absolute magnitude and morphological type. In gray are indicated the fields
		covered by the du Pont and Swope pointings listed in Table~\ref{table-Ha-observations}.		
\label{Dorado-group}}
\end{figure*}

The indication of SF and of dissipation events, reported for Dorado 
members in the current literature, motivated the present work.
In this  paper we aim at investigating the SF
of the Dorado members, both early and late-type galaxies, via \HaN\ narrow band imaging
observations.  Several studies used \Ha\ fluxes as a SF indicator and calibrated 
the relation \citep[see the review][]{Kennicutt1998}. As  a reference we mention the 
\Ha\ surveys  of LTGs  by \citet{James2004} and
\citet{Kennicutt2009}. \citet{James2004} found a strong correlation between total 
star formation rate (SFR hereafter) and Hubble type. The strongest SFR is 
found in isolated galaxies and occur in Sc and Sbc types.
The Dorado group has not been included in the
\citet{Kennicutt2008} study of the local volume. However, some
\Ha\ studies have been dedicated to NGC 1566, the brightest spiral member. 
Its \Ha\ total luminosity, $L_{H\alpha}$, has been
obtained by \citet{Hoopes2001} and \citet{Kennicutt2009}.
\citet{Roy1986} studied the \Ha\ emission in the outer NW arm of the galaxy. 
Uncalibrated \Ha\ observations of IC 2058, an edge-on spiral spiral member, 
have been performed by \citet{Rossa2003} with the aim of studying the
\Ha\ extra-planar distribution. \citet{Trinchieri1997} obtained a deep \HaN\
image of NGC 1553, the brightest among ETGs in the group.

\begin{center}
	\begin{table*}
		\centering
		\caption{H$\alpha$ narrow band observations  \label{table-Ha-observations}}%
		\tabcolsep=0pt%
		\begin{tabular*}{40pc}{@{\extracolsep\fill}lccccccr@{\extracolsep\fill}}
			\hline
			\textbf{Field} &  \textbf{Telescope} &  \textbf{Observing date} & \textbf{Exp. Time}  & \textbf{Filter}  & \textbf{Scale} & \textbf{seeing}  & Notes \\
			\textbf{ID }          &  \textbf{}    &  \textbf{}     & \textbf{[s]}      & \textbf{Ident.}  & \textbf{[arcsec px$^{-1}$]}  & \textbf{[arcsec]} &\\
			\hline
			NGC 1533 &  du Pont  & Nov. 30th/Dec. 1st,  2018  & 900$\times $3  &  716 & 0.259 &0.94$\pm$0.19& 5 frames, FoV=17\farcm7$^{2}$\\
			&                           &                              & 900$\times $3  &  657  & 0.259 & & 5 frames, FoV=17\farcm7$^{2}$\\
			NGC 1536   & du Pont  & Dec. 5th,  2018  & 900$\times $3  &  723 & 0.259 & 1.03$\pm$0.05&   1 frame FoV=8\farcm85$^{2}$\\
			&                           &                                    & 900$\times $3  &  657  & 0.259 &   & 1 frame FoV=8\farcm85$^{2}$\\
			NGC 1543 &  du Pont  & Dec. 3rd/5th,  2018  & 900$\times $3  &  723 & 0.259 & 0.96$\pm$0.09& 2 frames, FoV= 17\farcm7$\times$8\farcm85\\
			&                           &                                    & 900$\times $3  &  657  & 0.259 & & 2 frames, FoV= 17\farcm7$\times$8\farcm85\\
			NGC 1546   & du Pont  & Dec. 5th,  2018  & 900$\times $3  &  723 & 0.259 &1.26$\pm$0.05   & 1 frame FoV=8\farcm85$^{2}$\\
			&                           &                                    & 900$\times $3  &  657  & 0.259 & &  1 frame FoV=8\farcm85$^{2}$\\
			NGC 1549 &  du Pont  & Dec. 2nd,  2018  & 900$\times $3    &  716 & 0.259 &0.95$\pm$0.16 & 4 frames, FoV=  17\farcm7$^{2}$\\
			&                           &                                    & 900$\times $3    &  657  & 0.259 & & 4 frames, FoV=  17\farcm7$^{2}$\\
			NGC 1553 &  du Pont  & Dec. 2nd,  2018  & 900$\times $3    &  723 & 0.259 & 0.90$\pm$0.06&4 frames, FoV=  17\farcm7$^{2}$\\
			&                           &                                    & 900$\times $3    &  657  & 0.259 & & 4 frames, FoV=  17\farcm7$^{2}$\\
			IC 2058   & du Pont  & Dec. 3rd,  2018  & 900$\times $3 &  723 & 0.259 & 1.54$\pm$0.05 & 1 frame FoV=8\farcm85$^{2}$\\
			&                           &                                    & 900$\times $3  &  657  & 0.259 &  & 1 frame FoV=8\farcm85$^{2}$\\
			NGC 1566   & Swope & Dec. 6th,  2018  & 1200$\times $3  &  723 & 0.435 &1.60$\pm$0.03  & 1 frame FoV=29\farcm7$\times$29\farcm8\\
			&                           &                                    &1200$\times $3  &  657  & 0.435 &  &1 frame FoV=29\farcm7$\times$29\farcm8\\
			NGC 1581   & Swope & Dec. 7th,  2018  & 1200$\times $3  &  723 & 0.435 &  1.67$\pm$0.04&1 frame FoV=29\farcm7$\times$29\farcm8\\
			&                           &                                    & 1200$\times $3  &  657  & 0.435 & & 1 frame FoV=29\farcm7$\times$29\farcm8\\
			NGC 1596   & du Pont & Dec. 5th,  2018  & 900$\times $3  &  723 & 0.259 & 1.63$\pm$0.01&1 frame FoV=8\farcm85$^{2}$\\
			&                           &                                    & 900$\times $3  &  657  & 0.259 & & 1 frame FoV=8\farcm85$^{2}$\\
			\hline
		\end{tabular*}
		\tablefoot{Field ID refers to the main galaxy in the field. Col. 2 reports the LCO Telescope used. 
Col. 3 and Col. 4  report the observing date and the total reduced exposure time. 
Col.5 and Col. 6 quote  the off-on H$\alpha$ filters and the image scale. The FWHM of the filters H$\alpha$657, 716BP7 and 723BP10 are 77, 77, 109\AA, respectively (see also Figure~\ref{HA-filter}). 
In Col. 7 the average seeing measured in the frames reported in the Notes of Col. 8 which 
provides also the total fied of view covered. Images are obtained using as
{\tt GAIN}=1.54 e$^{-}$ ADU$^{-1}$ and  {\tt READOUT}=6.5 e$^{-}$  for the SITe2K CCD @ du Pont;  
{\tt GAIN}=1.040 e$^{-}$ ADU$^{-1}$ and {\tt READOUT}= 3.0 e$^{-}$  for the E2V CCD231-84 CCD @ Swope.}
	\end{table*}
\end{center}
\medskip

Our work is based on \HaN\ images  obtained at the Ir\'en\'ee du Pont 2.5m 
and Henrietta Swope 1m telescopes at Las Campanas Observatory (Chile).  
The paper plan is the following.  In \S~\ref{Dorado-description} we
describe the Dorado group and suggest it
 is still undergoing a strong transition phase.
 Appendix A provides the members list adapted from the dynamical
studies of \citet{Kourkchi2017} and \citet{Firth2006}. 
In \S~\ref{Observations} we present  \HaN\ observations and
the reduction performed. The analysis of the \HaN\ data is presented
in \S~\ref{Data-Analysis}.  In \S~\ref{Results} we discuss the morphology 
of the \HaN\ emission in each individual galaxy. The  comparison 
of our results with the literature is provided in 
Appendix~\ref{comparison_literature}.
The  SFR is discussed as a function of the members morphological type. 
In the context of group evolution, \S~\ref{Discussion} discusses 
the SFR  in the group sub-structures.  

\section{Dorado group: an overview}
\label{Dorado-description}

Before presenting our observations we describe the structure of the group, 
discussing also the membership of the galaxies which form its backbone,
i.e. our targets.

The Dorado group (RA=64.3718 [deg], Dec=-55.7811 [deg]) extends for 
several degrees in the Southern Hemisphere \citep[see e.g.][and reference
therein]{Firth2006,Kourkchi2017} (Figure~\ref{Dorado-group}). 
In this paper we assume that members are at the group
distance of 17.69 Mpc \citep{Kourkchi2017}.
The group includes both bright ETGs and LTGs.
Its structure is quite loose and clumpy so that the two giant galaxies  
NGC 1566, a grand-design spiral, and NGC 1553, a S0, have been considered 
the centre of two homonymous and independent groups. In this sense, different
member lists  have been compiled with different selection criteria.
The NGC 1566 group is composed of 18 member galaxies according to 
\citet{Huchra1982}, while 6 and 4 galaxies are members considered 
by \citet{Garcia1993} and \citet{Brough2006}, respectively. 
\citet{Makarov2011} reviewed the NGC 1553 group membership 
including 29 members. The Dorado faint galaxy population 
is still poorly defined. A large population of
dwarf galaxies is expected in groups dominated by ETGs, as is the case
of Dorado. From a CFHT {\tt Megacam} campaign, \citet{Tully2015} 
reported that these groups have a larger
dwarf/giant ratios with respect to groups dominated by LTGs
(considering giant M$_r < $ -17 and dwarf -17$ < M_r < $ -11, H$_0$ =
75 km~s$^{-1}$ Mpc$^{-1}$). An attempt of mapping the Dorado faint
galaxy population has been done by \citet{Carrasco2001} without,
however, determining their redshift.

\citet{Firth2006} redefined the Dorado members, on the
basis of the the group structure and its dynamical properties. The Firth
et al. Dorado backbone is composed of 20 members starting from the
79 galaxies, including few dwarfs that \citet{Ferguson1990} identified
and catalogued  in the Dorado region. The  \citet{Dressler1988} test
applied to the group candidates by \citet{Firth2006} provided a 2D view
of the group clumpiness. Furthermore, \citet{Firth2006}  performed a crossing-time
test checking for virialization and found indication that the group is
still undergoing dynamical relaxation. 

Recently \citet{Kourkchi2017} revised the group structure. 31 galaxies belong 
to Dorado according to their systemic radial velocities and membership
criteria of \citet{Tully2015}. 
The group velocity dispersion is 242 km~s$^{-1}$ \citep{Kourkchi2017}. The average
recession velocity is $\langle V_{hel} \rangle$=1230$\pm$89
km~s$^{-1}$. The 2$^\circ$  turnaround radius and the virial radius are 0.653 Mpc and
0.654 Mpc, respectively. The group virial mass weighted on virial radius
is 3.50$\times$10$^{13} $ M$_\odot$ (formul\ae\
and definitions are provided by \citet{Kourkchi2017} and
\citet{Tully2015}).

The list of candidate members adopted in this paper is presented
and discussed in Appendix~\ref{Dorado members}. We considered 31 member candidates,
30 from \citet{Kourkchi2017} (we exclude 2MASXJ04105983-5628496) 
and 1 from \citet{Firth2006} (PGC 75125). 
In Figure~\ref{Dorado-group} the distribution in
RA(J2000) and Dec (J2000) of the \citet{Kourkchi2017} members is shown.
In Appendix~\ref{Dorado members} we provide further information about the
group. Figure~\ref{fig_a1} shows the morphological type distribution
that includes  both ETGs and LTGs, and the velocity distribution.
The galaxy population is dominated by the two giant
ETGs, NGC 1549  and NGC 1553, the  pair AM 0414-554/AM 0415-555
\citep{Arp-Madore1987} located roughly in the projected group centre,
and the bright spiral, NGC 1566 at the group periphery.  
\citet{Iovino2002} suggested that the
pair NGC 1549/NGC 1553 is part of the SCG 0414-5559, a compact group
which includes also the spirals IC 2058 and the S0  NGC 1546.

NGC 1549 and NGC 1553, in the group projected centre, show a wide system 
of shells  and ripples \citep{Malin1983} whose origin is connected to
 merging/accretion episodes \citep{Dupraz1987,Weil1993,Mancillas2019}. 
While NGC 1553 has normal rotation curve, NGC 1549 shows a velocity gradient 
also along the minor axis \citep{Rampazzo1988}. 
NGC 1533  and NGC 1543, both S0s, show strong ring-like FUV emission
\citep{Marino2011a,Marino2011b}.   \citet{Rampazzo2017}
further investigated NGC 1533 and NGC 1543 using {\it SWIFT}-{\tt UVOT}
\citep{Roming2005} in W1, M2 and W2 filters confirming the presence of a young stellar
population. \citet{Mazzei2014a,Mazzei2014b} and \citet{Mazzei2019}
by matching the global properties of NGC 1533 and NGC 1543, derive their 
age and that of mergers (major and minor respectively) which drive
their evolution. \citet{Cattapan2019} 
deep surface photometry  unveiled strong interaction signatures 
in NGC 1533, IC~2038 and IC~2039.
NGC 1536 (AM 0409-563) is a late-type galaxy showing a clear asymmetry 
in its structure, likely the result of an on-going interaction.
NGC 1596 is lenticular galaxy interacting with NGC 1602, an LTG. In NGC
1596 the ionized  gas counter-rotate with respect to stars
\citep{Bureau2006}. \citet{Chung2006} suggested that the origin of the
counter-rotating gas is the interaction between the two galaxies
that are embedded in a common \HI\ cloud.

The literature investigated the presence of AGN like activity only in some of the
Dorado members. \citet{Annibali2010} found that NGC 1533 and NGC 1553
are LINERs in their nuclear region (they investigated the region r$_e$/8 - r$_e$/2,
but the classification, in their Table~4, is relative to the the r$_e$/16).
\citet{Rampazzo2013} studied  in the mid infrared with {\it Spitzer}
 the nuclear regions of NGC 1533, NGC 1549 and NGC 1553 unveiling PAHs, with
 anomalous ratios, in their spectra.  These features have been interpreted 
 as a relic of past star formation activity \citep[see e.g.][]{Vega2010}.

The high fraction of signatures of galaxy-galaxy encounters, accretions
and merging suggest that Dorado members are going through a deep transforming phase,
revealed by the (NUV - $r$) vs. M$_r$ CDM \citep{Cattapan2019}. Many ETGs 
have already reached the red sequence, while other galaxies are still transforming 
in the green valley. For the former, our objective is to test if their  star formation is
completely quenched. For galaxies in the green valley, we wish to check how
their SFR compares with that of galaxies of similar morphological type.

\begin{figure}
	{\includegraphics[width=8.9cm]{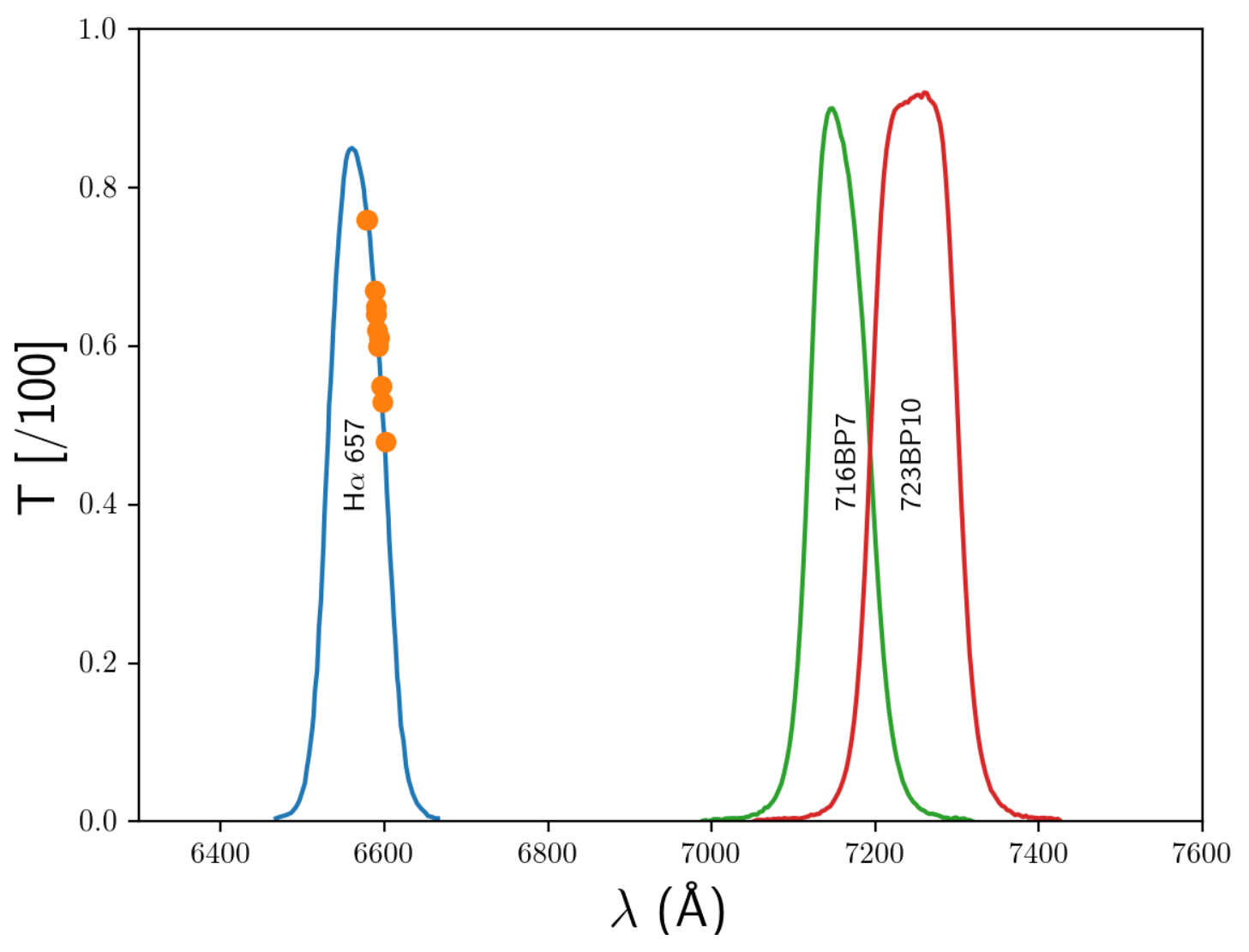}}
	\caption{Transmission profiles of the narrow band filters {\it on} and {\it off} the
		H$\alpha$ line. Observed galaxies are indicated as dots on the \HaN\
		filter at the wavelength given by their heliocentric recession velocity.
		\label{HA-filter}}
\end{figure}

\begin{figure*}[t]
	\center
	{\includegraphics[width=12cm]{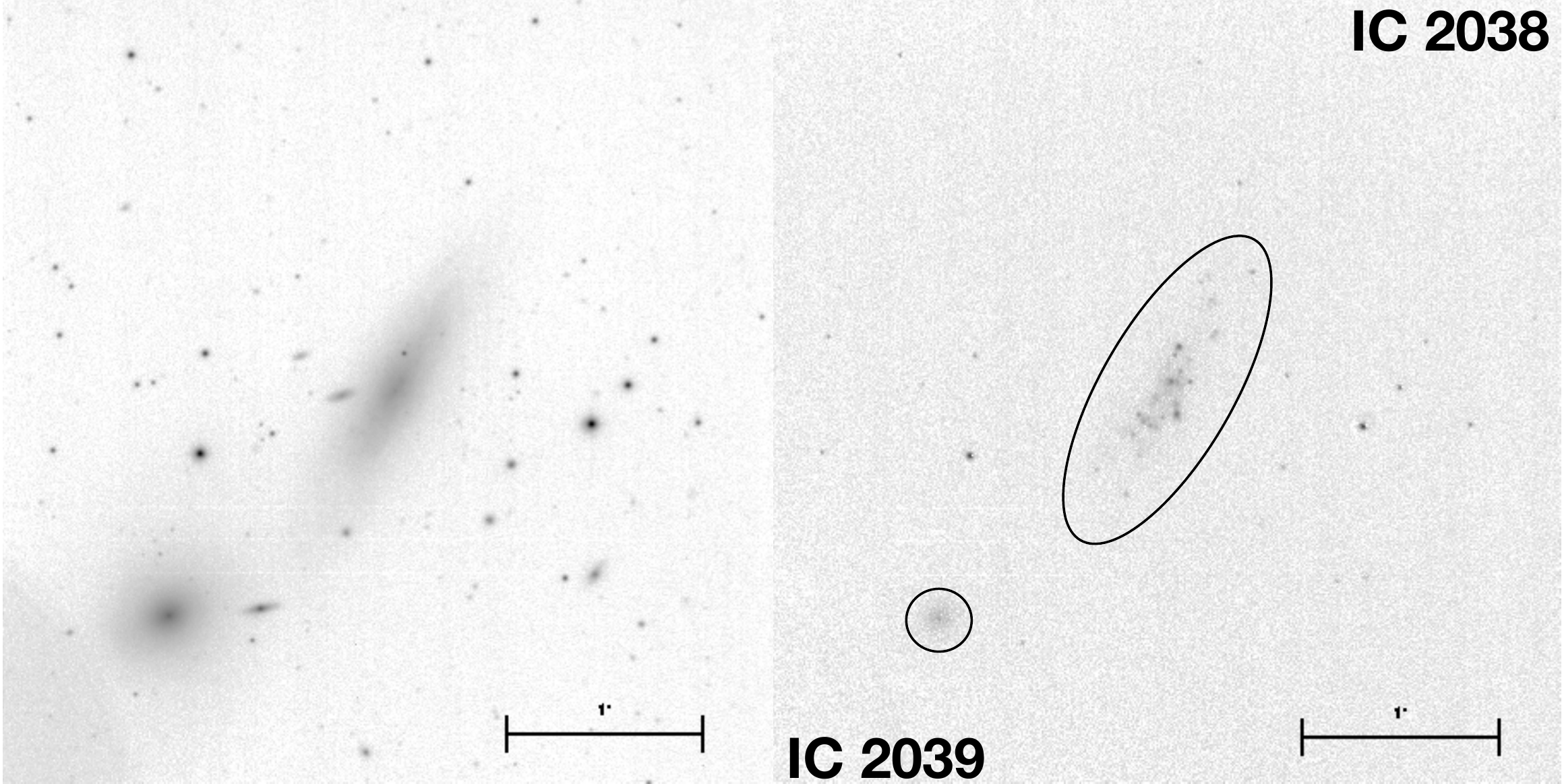}}
	{\includegraphics[width=12cm]{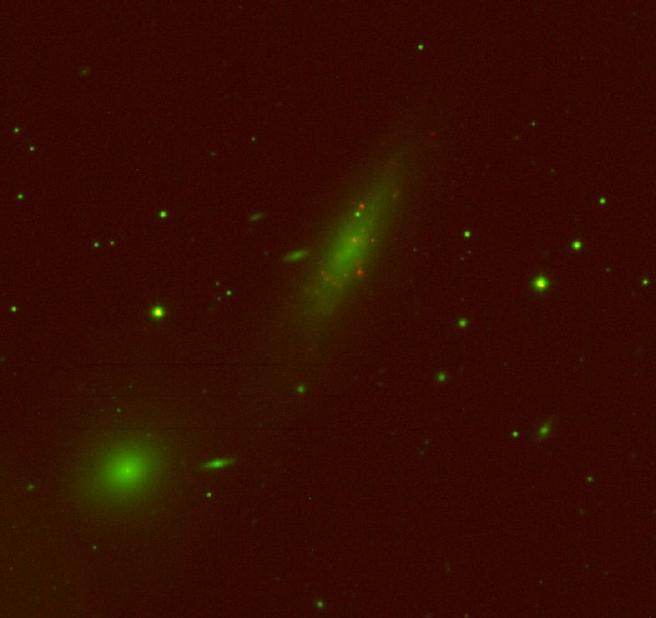}}
	\caption{The pair IC2038 (NW) and IC2039 (SE) in continuum 
		({\it top left panel}) and in the H$\alpha$+[NII] light ({\it top right panel}).
		The image size is  4\arcmin$\times$4\arcmin, North on the top East to the left.
		Residuals of stars have been masked.The ellipse, whose parameters 
		are given in Table~\ref{HA-measures}, encloses the integration area.
		The bottom panel shows a two-color image of the pair (continuum in green and 
		\HaN\ in red)  that highlights a large numbers of brights HII regions.} 
	\label{IC2038}
\end{figure*}

\begin{figure*}[t]
	\center
	{\includegraphics[width=12cm]{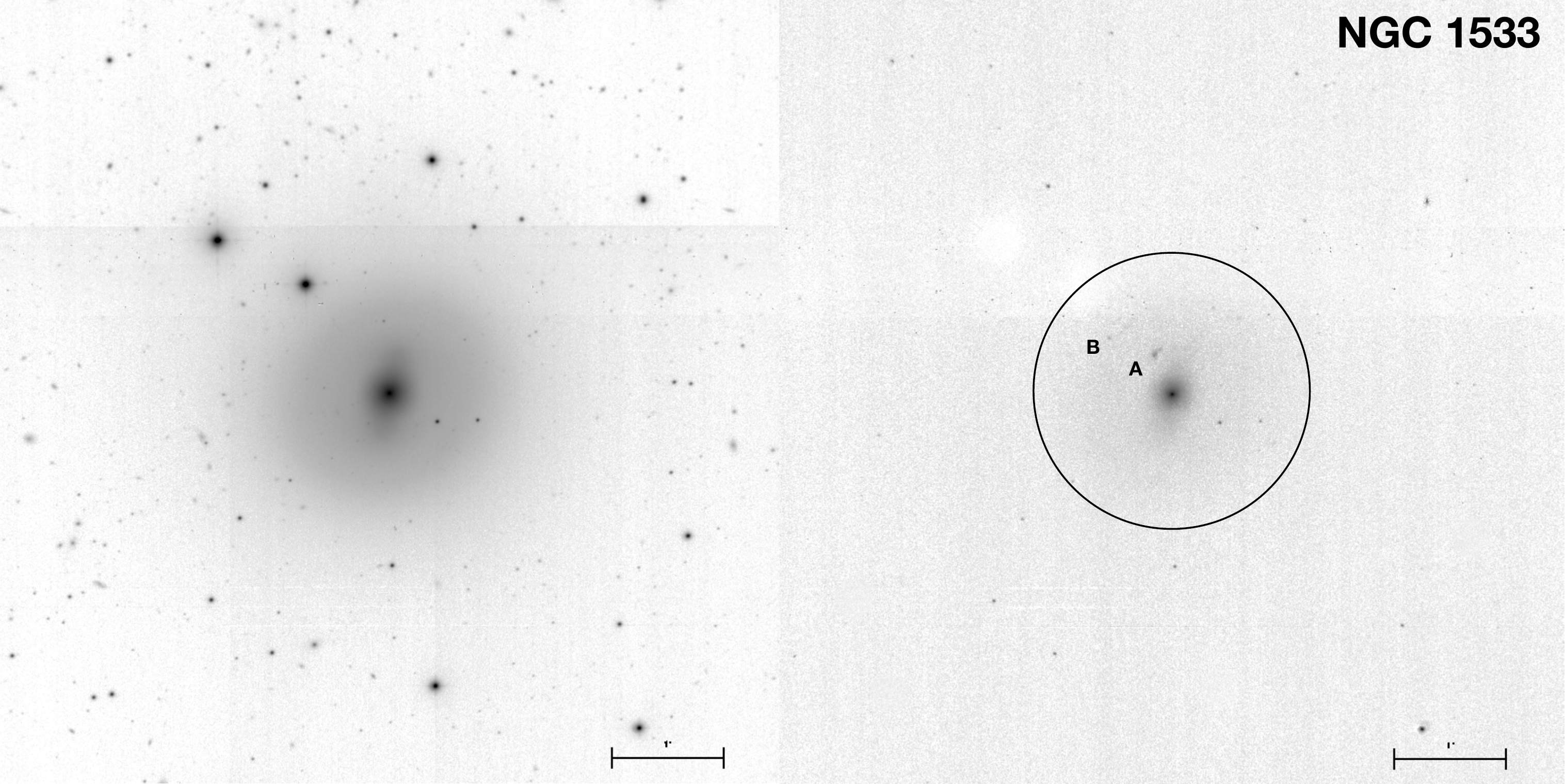}}
	{\includegraphics[width=12cm]{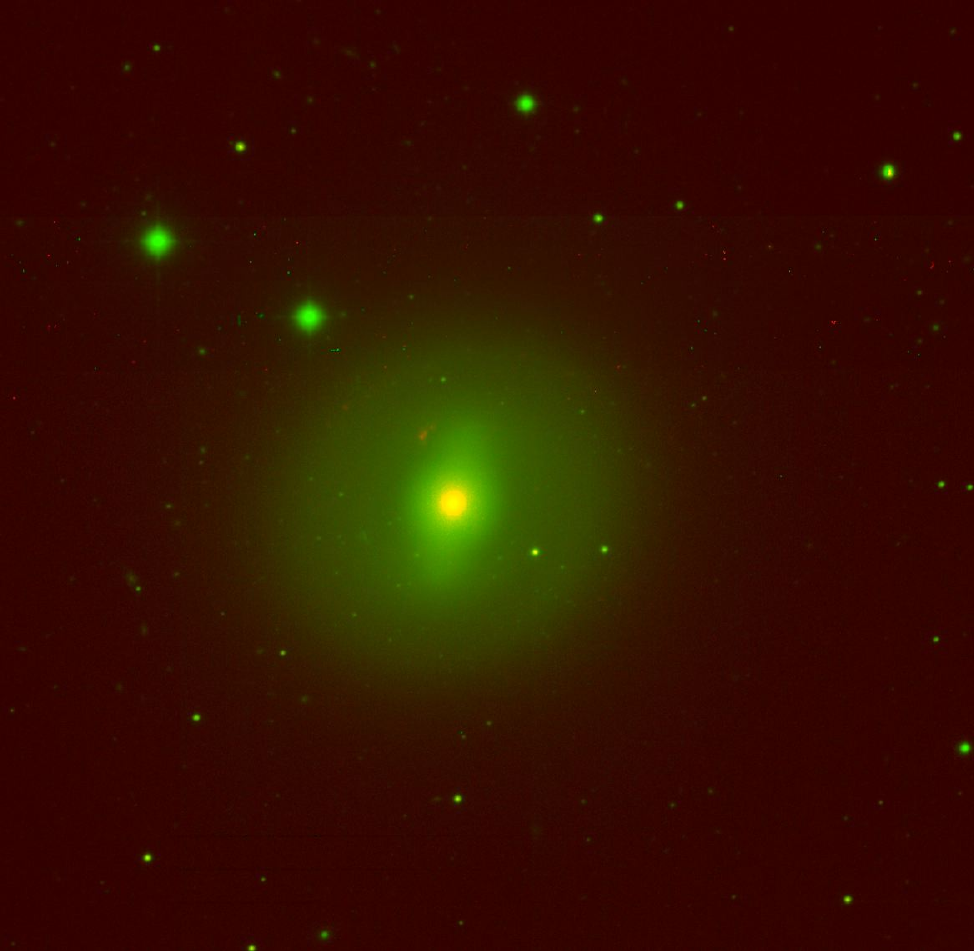}}
	\caption{As in Figure~\ref{IC2038} for NGC 1533. Labels A and B 
		indicate \HII\ complexes (see text). The image size is  
		7\arcmin$\times$7\arcmin.}
	\label{NGC1533}
\end{figure*}

\section{Observations and reduction}
\label{Observations}

In this section we describe our \HaN\ data set
and the reduction techniques we applied.  
The filter pass band includes [N II] emission 
lines (Figure~\ref{HA-filter}). The SFR of galaxies
is derived applying a correction to the \Ha\ emission,
following the recipe of \citet{Lee2009} (see Section~\ref{SFR}).

\medskip
A large fraction of Dorado members are dwarfs (see Appendix~\ref{Dorado_members}).
	The separation of most of them from other members would have
	required a specific pointing. We decided to concentrate our efforts on 
	bright galaxies for two reasons: 1)  they have nearby companions and 
	2) they are well studied galaxies, i.e. additional information about 
	their evolution (structure, kinematics, spectroscopic studies) are 
	already available in the literature. Considering that we have lost
	one and a half night for bad weather conditions, the final sample 
	resulted of 14 galaxies (13 in \citet{Kourkchi2017} plus PGC 75125
	in the \citet{Firth2006} list).

\subsection{\HaN\ observations}
\label{HA-Observations}

\HaN\ images have been obtained using the direct CCD Camera
at the du Pont 2.5m and at Swope 1m telescopes at Las
Campanas Observatory (LCO, Chile). Observations  
cover the period between November  30th and December 7th, 2018. 
Table~\ref{table-Ha-observations} reports the details of the observing runs.

The SITe2K CCD at the du Pont telescope covers a field of 
view (FOV) of 8.8 arcmin$^2$ with a spatial scale of 0\farcs259 
pix$^{-1}$, while the E2V CCD231-84 CCD at the  Swope telescope 
has a larger FOV of 29.7$\times$29.8 arcmin$^2$ with a spatial  
sampling of  0\farcs435 px$^{-1}$.
Table~\ref{table-Ha-observations} shows that some
du Pont observations are composed of a mosaic of frames for the bright
Dorado members, and a single  frame for less extended galaxies. 
NGC 1566 and NGC 1581 have been observed  with a single, wide field  
image at the Swope telescope.

\HaN\ and nearby continuum images have been obtained with
following filters: on-band \#657 (FWHM = 74 \AA ), and off-bands
\#716 (FWHM = 77 \AA ) and \#723 (FWHM = 109 \AA).
The \HaN\ \#657 contains the [N\, II]$\lambda\lambda$6548,6584
emission lines, while the \#716 filter can be contaminated by [Ar\,
III]$\lambda$7136 that is however a weak line, and \#723 is
almost completely free of emission lines.

The strategy of the observations consisted of taking 3$\times$900 sec
(at the du Pont telescope) and 3$\times$1200 sec (at the Swope telescope) 
images for each filter on-band and off-band.
The off-band image is obtained after each on-band image  
in order to minimize the seeing variations and to facilitate the
continuum subtraction. 
For some of the galaxies more than one pointing was required to cover 
the desired FOV. The seeing changed from night to night between
0\farcs9 and 1\farcs6, but it was rather stable within each single night.

\subsection{\HaN\ reduction}

Data reduction has been performed using the {\tt IRAF} reduction package 
\citep{Tody1986}. Raw scientific data have been corrected for overscan. 
Flat field correction was applied using twilight sky flats acquired during 
the observing run. Each image was background subtracted, normalized 
to 1 sec exposure time, and corrected for airmass by means of the CTIO
atmospheric extinction curve.

Each off-band image was first corrected for the different transmission
curves of  \#716 and \#723 with respect to the on band \Ha\ \#657 
filter, and then registered to the on-band image with the {\tt IRAF} task {\tt
IMALIGN}. In addition, the {\tt PSFMATCH} task was applied to take into
account the small seeing variations before proceeding with the
continuum subtraction. The continuum-subtracted, calibrated individual  
\HaN\ images have been combined to finally produce a single averaged 
image per pointing. Multiple pointings, when available, were combined 
to compose the mosaic using {\tt SWarp} \citep{Bertin2002}.

Some bright stars apart, the subtraction procedure was generally successful, 
since most of the stars disappeared with small or even no residuals. 
Saturated stars can not obviously be well subtracted and leave
large residuals on the final images.

\subsection{Flux calibration}

Photometric standard stars LTT 1788, LTT 3218, LTT 3864, LTT 4816, and
Feige 110, observed during the nights, have been used for flux
calibration. Raw images were processed as just explained up to airmass
correction. In order to calculate the flux calibration constants, the
{\tt PHOT} task was used to measure the instrumental fluxes. These
fluxes were then compared to the real fluxes obtained by multiplying the
spectrum of each standard star (extracted from {\tt IRAF} database) for the
transmission curve of the on-band \#657 filter.

\section{Data Analysis}
\label{Data-Analysis}

Before carrying out measurements, the background of each 
continuum subtracted image was inspected and carefully re-fitted and
re-subtracted with the {\tt IRAF} task {\tt IMSURFIT}, if necessary. This is a
delicate point, since even very small gradients can produce spurious
fluxes when large areas of the image are integrated. To estimate
the depth of the newly processed images, we measured the average
surface brightness standard deviation, it ranges between 3 and 5
$\times$ 10$^{-17}$ erg cm$^{-2}$ s$^{-1}$ arcsec$^{-2}$. 

We applied the {\tt IRAF} task {\tt ELLIPSE} to the off-band images keeping fixed
center, ellipticity, and position angle in order to obtain concentric
apertures. The ellipse parameters, reported in Table~\ref{HA-measures} 
(columns 3,4 and 5)
have been chosen on the basis of the outer isophotes of the images in the off-band
continuum. Residuals of  stars, particularly bright and saturated stars,
have been carefully masked.  Figure~\ref{IC2038} and Figure~\ref{NGC1533}
show on-band and off-band images of the  triplet \citep{Cattapan2019} 
composed of the spiral IC 2038 and of the two early-type galaxies IC 2039 
and NGC 1533. 

Once fitted the off-band images, the same ellipse has been applied to
the on-band continuum-subtracted ones. Following the examples and
suggestions by James et al. (2004) we calculated the \HaN\
equivalent width (EW) profiles, and the total flux within the 
largest apertures showing emission up to 1 $\sigma$  level. 
Finally, the fluxes have been corrected for the 
transmission of the on-band H$\alpha$ filter at the wavelength 
corresponding to the radial velocity of each single galaxy, 
as given in Table~\ref{Dorado_members}.
The 3$\sigma$ error  associated to each measure is calculated 
using the sky background measured in 5 independent regions 
outside the ellipse whose parameters are reported in
Table~\ref{HA-measures}. 

We remind the reader that fluxes are not internal reddening corrected since the
value of the extinction is unknown. Nonetheless it is worth to note
that reddening and [N\, II] contamination change the \Ha\ flux in
opposite directions, such that they can even compensate each other
but both effects cannot be quantified with the present data. 
This uncertainty does not apply to the EW, that is virtually reddening 
independent. The final \HaN\ fluxes and EW are reported in 
Table~\ref{HA-measures}.

\begin{figure*}
\centering
{\includegraphics[width=16.5cm]{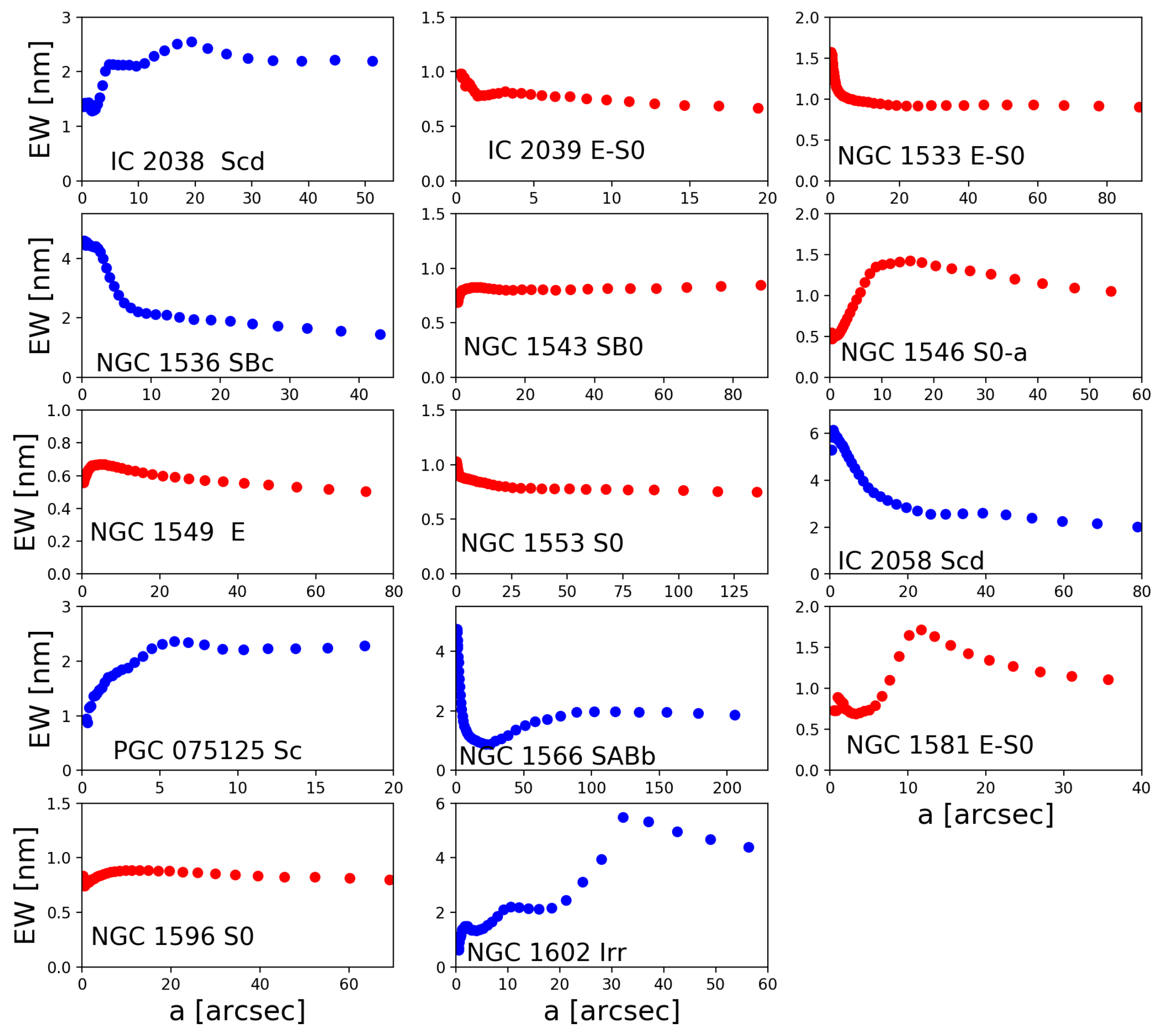}}
\caption{ The  H$\alpha$+[NII] EW growth curve  of each galaxy  measured in 
concentric elliptical apertures out to the value of the elliptical  aperture 
given in Table~\ref{HA-measures} and shown
in Figure~\ref{IC2038}, Figure~\ref{NGC1533} and in Appendix B (Figure~\ref{NGC1536}
-- Figure~\ref{NGC1596}). ETGs are indicated in red, LTGs
in blue. Morphological classifications are from {\tt HyperLeda}. 
  \label{EW}}
\end{figure*}

\begin{figure*}
\centering
{\includegraphics[width=16.5cm]{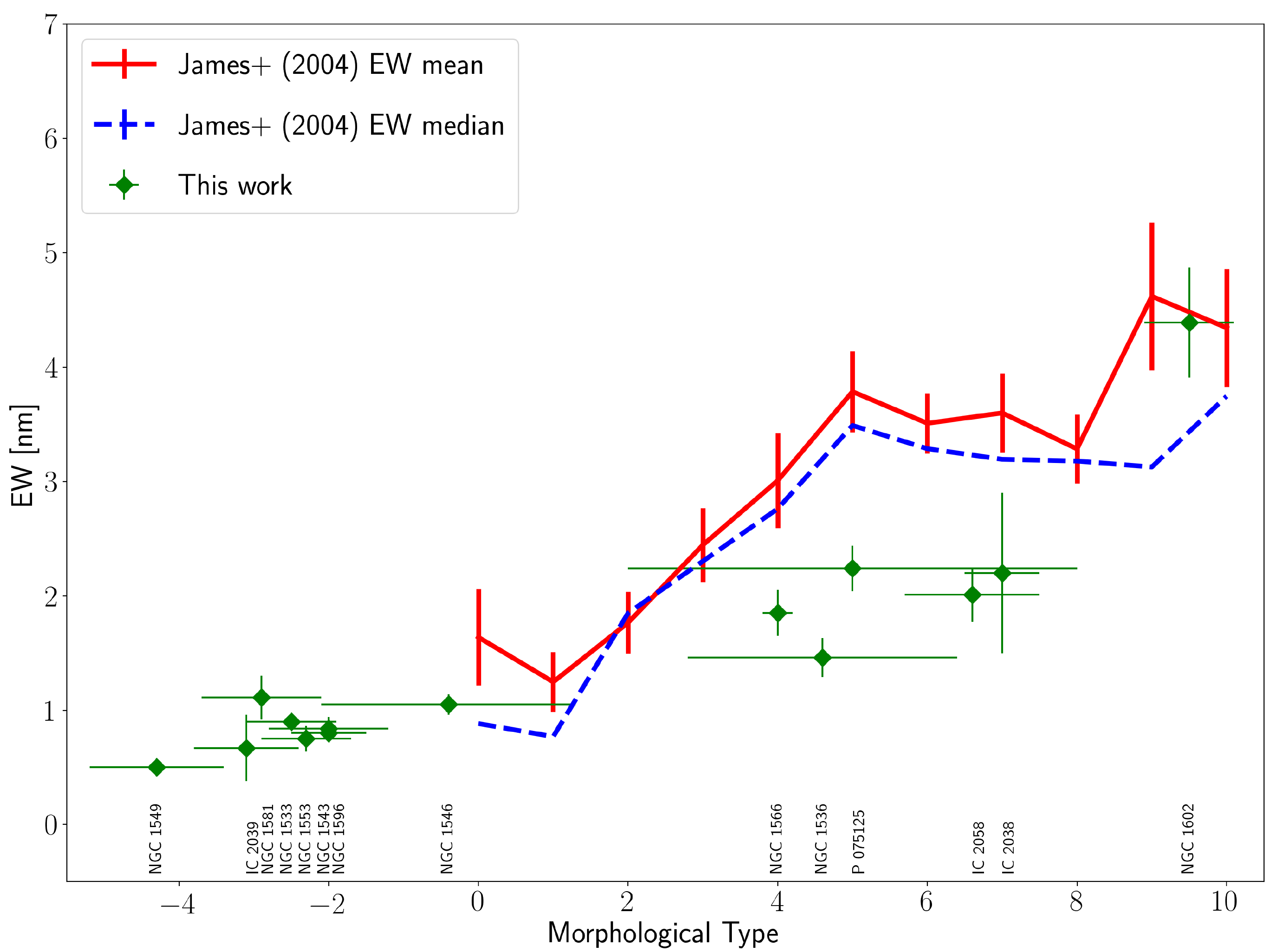}}
\caption{The integrated value  of H$\alpha$+[NII] EW for our targets as a 
function of their morphological type.  Our values are compared with the mean and 
the median for the late-type galaxy sample studied by  \citet{James2004}.
\label{EW_comparison}}
\end{figure*}

\begin{figure*}[t]
\center
{\includegraphics[width=16.5cm]{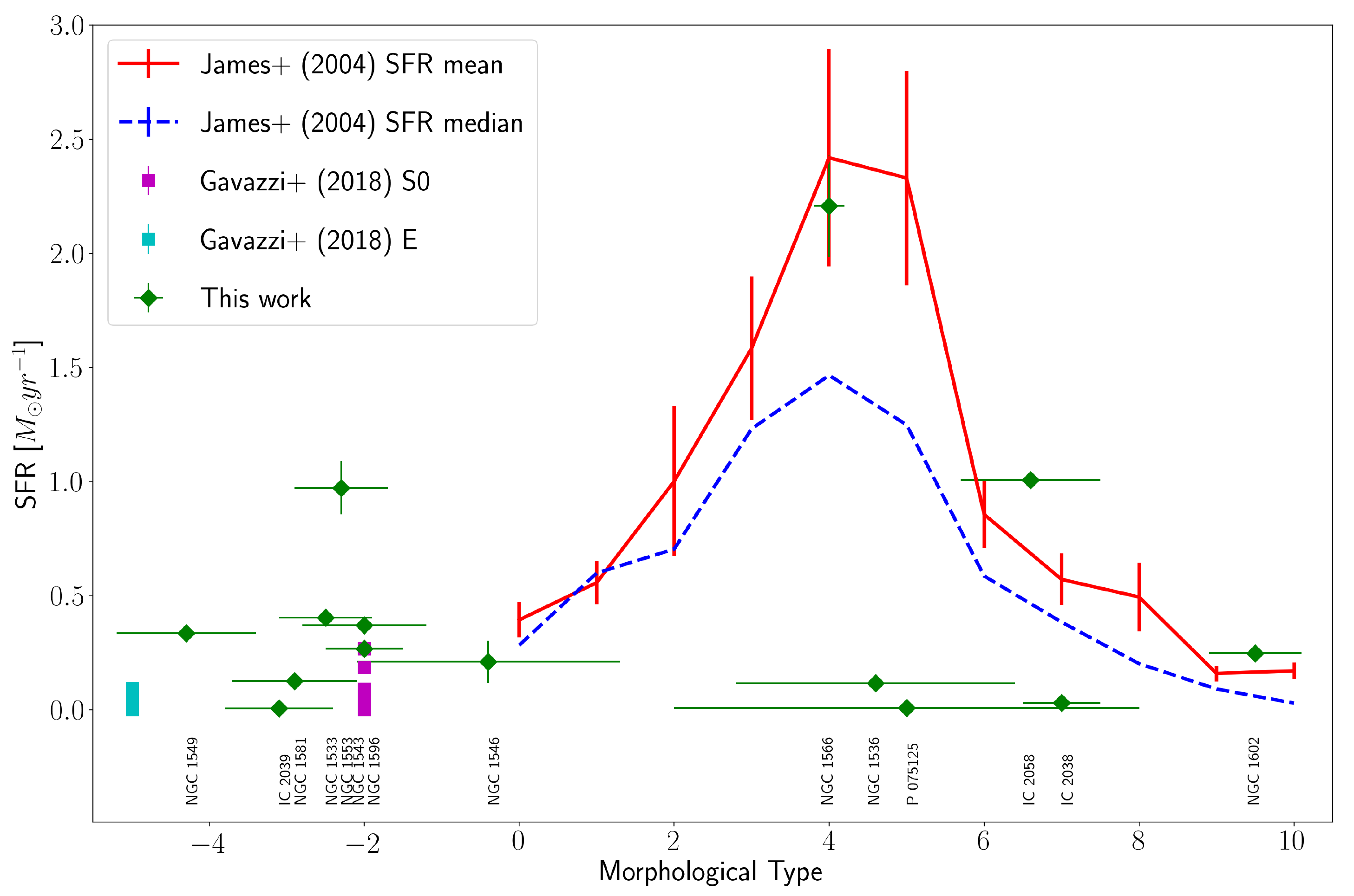}}
  \caption{H$\alpha$ SFR of  Dorado galaxies as a
  function of their morphological type. For LTGs (0$\leq$Type$\leq$10) the red
  line and the blue dotted line represent the mean and the median 
SFR computed on a sample of 334  galaxies by \citet{James2004}.
For ETGs ($-5\leq$Type$<$0) the magenta and cyan squares represent the 
SFR of detected objects (55/147), representative of the whole ATLAS$^{3D}$ 
survey,  by \citet{Gavazzi2018}.  These authors divided ETGs into the two families
of elliptical (Type=$-5$) and S0s (Type=$-2$). Our values  (green diamonds) are reported 
in Table~\ref{HAL-SFR}.  
  \label{SFR_comparison}}
\end{figure*}

\begin{table*}
\centering
\caption{The parameters used to measure the H$\alpha$+[NII] flux and its Equivalent Width
	\label{HA-measures}}%
\begin{tabular*}{40pc}{@{\extracolsep\fill}lccccc@{\extracolsep\fill}}%
	\hline
	\textbf{ID}        & \textbf{$\epsilon$} & \textbf{PA}            & a      &\textbf{F(H$\alpha$+NII)}  & \textbf{EW}\\
	\textbf{source}    &                     & \textbf{[$^\circ$] } & \textbf{[arcsec]} & \textbf{[$\times$10$^{-14}$ erg cm$^{-2}$ s$^{-1}$]} & \textbf{[nm]}\\
	\hline	
	IC 2038	   &    0.63 & 150	 & 51.3	 &  11.0$\pm$2.9   & 2.2$\pm$0.7\\
	IC 2039	   &    0.05 &   0	 & 19.4	 &   2.8$\pm$1.1   & 0.7$\pm$0.3\\
	NGC 1533   &    0.05 &   0	 & 89.2	 &  182.5$\pm$7.0  & 0.9$\pm$0.1\\
	NGC 1536   &    0.27 & 165	 & 43.0	 &   45.4$\pm$1.6  & 1.5$\pm$0.2\\
	NGC 1543   &    0.05 &   0	 & 88.0	 &  172.9$\pm$17.7& 0.8$\pm$0.1\\
	NGC 1546   &    0.19 & 145	 & 54.3  &   93.1$\pm$4.0   &1.0$\pm$0.1\\
	NGC 1549   &	0.05 &   0	 & 73.0	 &  192.7$\pm$10.5 & 0.5$\pm$0.1\\
	NGC 1553   &    0.25 & 145	 & 135.6 &  545.1$\pm$65.2 & 0.7$\pm$0.1\\
	IC 2058	   &    0.75 &  18	 &  79.1 &	33.3$\pm$3.6    &2.0$\pm$0.2\\
	PGC 75125  &    0.36 &  20   &  15.8 &   3.3$\pm$0.3   & 2.2$\pm$0.2\\
	NGC 1566   &    0.10 &  20   & 227.0 & 1246.7$\pm$125.1& 1.8$\pm$0.2\\
	NGC 1581   &    0.50 &  82   & 35.7 &	50.3$\pm$4.3   & 1.1$\pm$0.2\\
	NGC 1596   &    0.61 &  20   &  69.4 & 119.7$\pm$6.5   & 0.8$\pm$0.1\\ 
	NGC 1602  &     0.26 &  85   &  56.5 &	97.6$\pm$6.9   & 4.4$\pm$0.5 \\
\hline
\end{tabular*}
\tablefoot{Notes: Col.1 provides the source ident. Col.s 2,3,4 are  the ellipticity, 
$\epsilon$, the Position Angle (PA measured North to East) and the semi-major axis, 
$a$ of the ellipse within which the fluxes have been measured, respectively. 
Col. 5 is  \HaN\ flux measured within the elliptical 
aperture and the 3$\sigma$ error.  Col.6 is the EW within the ellipse.}
\end{table*}	

\section{Results}\label{Results}

	We now present the results of our analysis for each single galaxy.
	In Appendix~\ref{comparison_literature} we instead compare our results
	with the available literature.

\subsection{Continuum vs. \HaN\ morphology: individual notes}

We report below some notes about the \HaN\ emission in our targets 
with particular attention to asymmetries or correlations with 
features present in the continuum image and to  
resonance structures. We present and discuss two-color images 
(continuum in green and Ha+[NII] in red) which emphasize morphology 
of \Ha\ emission regions.

\medskip
\noindent
\underbar{IC 2038/IC 2039}~~~~~ 
The two galaxies are classified Scd (Type=7.0$\pm$0.5) and E-S0 (Type=-3.1$\pm$0.7),
respectively. According to \citet{Cattapan2019}  the
galaxies show clear signatures of interaction. The \Ha\ distribution of IC 2038, 
shown in Figure~\ref{IC2038}, is clumpy and marks the presence of \HII\ regions, 
without any obvious arm structure. \HII\ regions are distributed along the galaxy 
body which appears slightly elongated towards the companion galaxy, in the SE direction
\citep[see][their Figure 2]{Cattapan2019}.
 In IC 2039 the \HaN\ emission is diffuse and smooth, without \HII\ regions. 

\medskip
\noindent
\underbar{NGC 1533}~~~~~ The galaxy is classified E-S0 with bar (Type=-2.5$\pm$0.6) by 
{\tt HyperLeda} and (RL)SB0$^0$ by \citet{Comeron2014},
i.e. the galaxy has an outer ring, a lens and an inner bar. 
 \HII\ regions in Figure~\ref{NGC1533}, labeled A and
B, correspond to the rings clearly revealed in FUV by {\it GALEX} 
\citep[see e.g.][]{Werk2010,Marino2011,Rampazzo2017}.

\medskip
\noindent
\underbar{NGC 1536}~~~~~ The galaxy classification is  SBc (Type=4.6$\pm$1.8)
in {\tt HyperLeda}. In the continuum image the galaxy shows a thin bar in the central region 
and   irregular arms with  signatures of interaction. The \HaN\ emission follows
roughly the structures seen in the continuum (Figure~\ref{NGC1536}).
 \HII\ regions are asymmetrically distributed with respect to the inner bar. 
These regions develop  along the NW arm and are scattered within the 
South galaxy body.

\medskip
\noindent
\underbar{NGC 1543}~~~~~ The galaxy is considered a S0 with bar and a ring 
(Type=-2.0$\pm$0.8) in {\tt HyperLeda}. The \HaN\ emission follows the bar in 
the galaxy central region (see Figure~\ref{NGC1543}).
The \Ha\ emission is concentrated in some \HII\ complexes, labeled A, B and C 
in the figure, along the South East edge of the outer ring. 
These star formation areas  are clearly visible in FUV {\tt GALEX} observations 
\citep{Marino2011a,Marino2011b,Rampazzo2017}.

\medskip
\noindent
\underbar{NGC 1546}~~~~~ This galaxy, considered a S0-a (Type=-0.4$\pm$1.7) in  
\citet{Comeron2014} is classified  E(b)3/(R')SA(r)ab, i.e. can be a 3D ETG or a SA 
with an inner ring. 
The continuum image in Figure~\ref{NGC1546} shows a small bulge embedded
 in a strongly  inclined disk (53.3 degree in {\tt Hyperleda}). The \HaN\ emission shows a 
flocculent structure inside the disk. The \HII\ regions, which are emphasized in the 
two-color image, follow the inner and outer rings evidenced by \citet{Comeron2014}. 
A strong dust-lane is visible in the SW ridge of the galaxy. 

\medskip
\noindent
\underbar{NGC 1549}~~~~~ The galaxy is considered in {\tt HyperLeda} 
a true E  (Type=-4.3$\pm$0.9). 
The \HaN\ emission in Figure~\ref{NGC1549} (top right panel) shows an amorphous  
distribution like that of the continuum (top left panel). No \HII\ regions are visible.

\medskip
\noindent
\underbar{NGC 1553}~~~~~ The galaxy is classified S0 with ring (Type=-2.3$\pm$0.6). 
\citet{Comeron2014} classified it SA(rl,nr'l)0$^+$, i.e. two rings are present in
this lenticular galaxy.  75\% of the \HaN\ flux is measured within 
an ellipse of 60\arcsec\ of semi-major axis, containing the core and the ring 
visible in Figure~\ref{NGC1553}, i.e. in a relatively small central region with respect to
the size of the ellipse reported in Table~\ref{HA-measures}.
No \HII\ regions are revealed on the galaxy (see, however, $\S$~\ref{Dorado-compact}).
\citet{Trinchieri1997} show a \HaN\ image of the galaxy (their Figure~11, right panel).  The flux
is not provided in the paper.  At faint levels (they smoothed the image to enhance features), 
the morphology shows a noisy spiral arm, in  the South East direction, not detected
in our observations.

\medskip
\noindent
\underbar{IC 2058}~~~~~ The galaxy is an edge-on spiral  (Scd, Type=6.6$\pm$0.9). \HaN\ emission
 is revealed along the plane of the galaxy and no obvious bulge is visible (see Figure~\ref{IC2058}).
Uncalibrated \Ha\ observations of this galaxy, performed by \citet{Rossa2003}, 
did not detect any extra-planar \Ha\ emission  in the halo of this spiral galaxy. 
Figure~\ref{IC2058} shows that \HII\ regions are indeed distributed all along its disk. 

\medskip
\noindent
\underbar{PGC 75125}~~~~~ This dwarf galaxy 
is a companion of IC 2058 according to \citet{Firth2006}.
They are nearby in projection and have similar heliocentric velocity, $\Delta V_{hel}$=26 \kms. 
According to {\tt HyperLeda} the galaxy morphological type is T=5.0$\pm$3.0.
In the continuum image the galaxy structure appears boxy and without spiral arms. 
An elongated, distorted  \HaN\ distribution (Figure~\ref{PGC075125})  appears in the 
outer regions of this galaxy.  The \HaN\ emission could mark the (warped) disk of the galaxy. 
We suggest this galaxy is likely a late S0.

\medskip
\noindent
\underbar{NGC 1566}~~~~~ 
This is a grand design, barred galaxy, classified as a SABb (Type=4.0$\pm$0.2) 
in {\tt HyperLeda} catalog. According  to \citet{Comeron2014}, who classified 
the galaxy  (R'1)SB(r'l, s,nb)b, there are two rings of which  we do not find any 
evidence in the \HaN\ image (Figure~\ref{NGC1566}). NGC 1566 is
asymmetric (the Southern part is more stretched than the Northern one).

\medskip
\noindent
\underbar{NGC 1581}~~~~~ 
This is an E-S0 galaxy (Type=-2.9$\pm$0.8 according {\tt HyperLeda}). 
Figure~\ref{NGC1581} shows an inner ring both in the continuum and in the \HaN\  images. 
 Small, bright \HII\ regions are visible near the galaxy centre and on the inner ring, which 
 is  brighter in the west region of the galaxy. NGC 1581 is  likely a companion of NGC 1566
 \citep{Kendall2015,Oh2015} (see $\S$~\ref{Discussion}).

\medskip
\noindent
\underbar{NGC 1596/ NGC 1602}~~~~~ These two Dorado
members, nearby in projection, are separated by $\Delta V_{hel}$=230 \kms\ 
(Figure~\ref{NGC1596}).  NGC 1596 is a  S0 galaxy (Type=-2.0$\pm$0.5). No \HII\ regions are visible. 
NGC 1602 is an Irregular (Type=9.5$\pm$0.6). Our \HaN\ image outlines the presence of several  
\HII\ regions which form a sort of ring and a large complex, west of the galaxy centre.

\begin{center}
\begin{table*}[t]%
\centering
\caption{Parameters used to compute from H$\alpha$ luminosity the SFR 
\label{HAL-SFR}}%
\begin{tabular*}{40pc}{@{\extracolsep\fill}lcccc@{\extracolsep\fill}}%
\hline
\textbf{ID}    & \textbf{M$_B$} & \textbf{E(B-V)} &\textbf{ L(H$\alpha$)}  & \textbf{SFR$_{H\alpha}$}    \\
\textbf{source}&                &                 &\textbf{[10$^{33}$ W ]} &\textbf{[M$_\odot$ yr$^{-1}$]} \\
\hline
IC 2038  &  -15.87 &  0.010  & 0.4$\pm$0.1 & 0.03$\pm$0.01  \\
IC 2039  &  -16.30 &  0.010  & 0.009$\pm$0.004 & 0.008$\pm$0.003  \\     
NGC 1533 & -19.52  & 0.016   & 4.9$\pm$0.2 & 0.40$\pm$0.02 \\
NGC 1536 & -17.80  & 0.019   & 1.42$\pm$0.05 & 0.118$\pm$0.004 \\
NGC 1543 & -19.88  & 0.024   & 4.4$\pm$0.5 & 0.37$\pm$0.04 \\
NGC 1546 & -19.24  & 0.013   & 2.6$\pm$0.1 & 0.21$\pm$0.09 \\
NGC 1549 & -20.61  & 0.011   & 4.2$\pm$0.2 & 0.4$\pm$0.02 \\
NGC 1553 & -21.02  & 0.013   &11.9$\pm$1.4 & 1.0$\pm$0.1  \\
IC 2058  & -17.55  & 0.014   & 1.1$\pm$0.1 & 1.01$\pm$0.01  \\
PGC 75125& -15.81 & 0.014   & 0.11$\pm$0.01 & 0.01$\pm$0.01 \\
NGC 1566  &-20.99  & 0.008  & 27.2$\pm$2.7 & 2.2$\pm$0.2 \\
NGC 1581 & -17.71  & 0.007   & 1.6$\pm$0.1 & 0.13$\pm$0.01 \\
NGC 1596  &-19.27  & 0.008   & 3.3$\pm$0.2 & 0.27$\pm$0.02 \\
NGC 1602 &-17.87   & 0.009   & 3.0$\pm$0.2 & 0.25$\pm$0.02 \\

\hline
\end{tabular*}
\tablefoot{Parameters used to compute the H$\alpha$ fluxes and the galaxy SFR.
Col.1 source ident. Col. 2 absolute B-band magnitude (17.69 Mpc from
\citet {Kourkchi2017}). Col.3 E(B-V) extinction.
Col. 4  and col. 5 report the \Ha\ luminosity and the SFR$_{H\alpha}$. 
The SFR has been computed from the H$\alpha$ luminosity according to \citet{Lee2009}
(see text in \S~\ref{SFR}).}
\end{table*}
\end{center}

\subsection{\HaN\ equivalent width} 

In Figure~\ref{EW} we plot the EW growth curve for each galaxy.  Growth curves, 
computed following \citet{James2004}, whose large sample is a benchmark reference,
are obtained dividing the \HaN\ emission aperture fluxes by the continuum within
the same elliptical aperture, assuming that the average continuum level within the filter 
is equal to the continuum level in \Ha\ \#657 filter.  Total fluxes and EW are reported
in Table~\ref{HA-measures}. The EW of the \HaN\ emission traces the specific SFR,
being normalised by the luminosity of the older stellar population of the galaxy \citep[see
e.g.][]{James2004,Lee2009}.

Figure~\ref{EW} shows that EW growth curves are
basically flat in ETGs with average values $\leq0.75$ nm. 
NGC 1553, IC 2039 and NGC 1533 show higher EW values at the centre, 1, 1, and 1.5,
respectively. Moreover, the quasi-constant trend of  NGC 1533 sets on an higher 
value than on average.  NGC 1546 shows a rising inner curve 
reaching  EW=1.5, between 10\arcsec\ and 20\arcsec, that is in the bright ring  
and then a slow decrease, down to EW=1  in outer region (about 1\arcmin).

NGC1581 shows almost the same trend as NGC 1546, reaching EW=1.7 at 12" 
and then slowly decreasing to EW=1
at 35\arcsec, the limit of emission we detect in this galaxy.

In LTGs growth curves show different shapes. IC 2038 shows a rising 
curve out to 18\arcsec\ reaching a value of 2.5 units then decreasing and  
stabilising around EW=2. In PGC 75125 the curve shows a plateau following the 
maximum value, EW=2.5 reached at 5\arcsec. Growth 
curves of NGC 1536, IC 2058 are decreasing from the centre to the periphery, 
while in NGC 1566, after reaching a minimum around 25\arcsec, the curve 
shows a slight increase.

Concerning Irregular galaxies, NGC 1602 shows the higher value in our sample, 
EW=5.6  at 30\arcsec.  In the outer region the curve decreases down to EW=4.2.
 
 The total EW  (Table~\ref{HA-measures})  of our targets is plotted versus their
 morphological type in Figure~\ref{EW_comparison}. In the same figure we compare 
 our results with those  of  \citet{James2004} which report the mean EW as derived 
 from a large sample of Spiral galaxies, including interacting spirals as well as 
 members of pairs and groups.
They found an increasing trend of EW.  Our results follow this trend although our targets 
 with 4$\leq$Type$\leq$8 have values lower than the average and the
 median. However, spirals with values of EW$\approx$1.5--2.2, as in our
 Dorado members, are present in the \citet{James2004} sample (see their Figure 15).
 
Although our sample  does not allow any statistical significance, with respect to previous studies
it extends the range of morphological types including several ETGs (the brigher ones) 
in the Dorado groups.
  
\subsection{Star Formation Rate}
\label{SFR}

The \Ha\ luminosities and SFRs are reported in the last columns of Table~\ref{HAL-SFR}. 
Since our observations include [N II] we need to remove
this contribution from the total flux. This implies the knowledge 
of the [N II]/\Ha\ line ratio, e.g. from spectroscopy. 
The alternative is to use  the average relationship
between [N II]$\lambda$6583/\Ha\ and the total B-band absolute magnitude
M$_B$ (col. 2 in Table~\ref{HAL-SFR}),  a consequence of the 
luminosity-metallicity relation for galaxies.
We adopt the \citet{Lee2009} formul\ae, in particular their equations 1 and 2, 
to derive the [N II]/\Ha\ ratio, L(\Ha) and the SFR.
Assuming 3 to 1 ratio between [N II]$\lambda$6583 
and [N II]$\lambda$6548, they propose the following formul\ae. If M$_B>$-20.3,
\noindent
log([N II]$\lambda$6583/\Ha )=(-0.173$\pm$0.007)$\times$M$_B$-(3.903$\pm$0.137).
\noindent
For  M$_B$ $\leq$-20.3, [N II]$\lambda$6583/\Ha=0.54.
For all LTGs, with the exclusion of NGC 1566, since they are fainter than -18 
absolute B-mag, [N II]/\Ha\ $\leq$0.16 i.e. an error $<$10\%. 
In this way we compute the \Ha\ luminosity, L(\Ha),  in Table~\ref{HAL-SFR}
and the SFR as: SFR(M$_{\odot}$~yr${^-1}$)=7.9$\times$10$^{-35}$  L(\Ha) W.

In Figure~\ref{SFR_comparison} we plot the SFR as function of the
morphological type. For Type $>$ 0, three of our targets are consistent with  the trend reported by 
\citet{James2004} (for a sample of 334 spiral galaxies, of morphological
types from S0/a to Irr and with V$_{hel}\leq3000$ \kms), increasing until Type $\sim$4--5 
then declining, while other three targets, PGC 75125, IC 2038 and NGC 1536  have 
significantly lower  SFR, comparable with Type$<$0 sample galaxies. 
We point out that LTGs with low SFR,  like the cases above, are also found in the 
\citet{James2004} sample. PGC 75125, in particular, has a very
uncertain classification, and is likely near to S0s  (see our Section~\ref{Results}).

\bigskip
Dorado is not rich in spirals enough  to define a trend for Types$>0$,  
but its galaxy population includes several early-types. 
In Figure~\ref{SFR_comparison} we compare our
results with the SFR calculated by \citet{Gavazzi2018} from 
\HaN\  imaging of a sample of 147 ETGs from ATLAS$^{3D}$. 
They detect 55 EGTs (37\% of the sample), mostly S0.
To compute the SFR \citet{Gavazzi2018} adopted
the \citet{Kennicutt1998} formula modified by a Chabrier IMF.
Since SFR(Chabrier) = SFR(Kennicutt)/1.58 \citep{Boselli2015} 
we applied this transformation to the sample 
in \citet{Gavazzi2018} to compare their values  with our estimates. 

\citet{Gavazzi2018}  divided  ETGs in two sets: Es (Type=$-5$, magenta squares) 
and S0s (Type=$-2$, cyan squares). Four of our galaxies show a larger SFR than 
\citet{Gavazzi2018} sample, 3 S0s (NGC 1533, NGC 1543, NGC1553)
and NGC 1549, a bona fide E.  NGC 1553 has a SFR remarkably above 
the average of ETGs, despite its  EW value. 
Figure~\ref{NGC1553_FP} \citep[adapted from][]{Rampazzo2003},
discussed in \S~\ref{Dorado-compact}, provides a clear map of the 
presence and of the complex structure of the \Ha\ emission in 
the inner region of this galaxy.

\section{Discussion}
\label{Discussion}

It is widely accepted that cluster galaxies have depressed  star formation 
rates (SFRs) in comparison  with the field 
\citep[e.g.][and references therein]{Poggianti2006,Vulcani2010,Paccagnella2016}. 
The understanding of the SF quenching during the galaxy co-evolution in groups
is fundamental for two reasons:
groups contain $\approx$60\% of the galaxies in the nearby universe 
\citep{Tully1988,Ramella2002,Tago2008} and the transition between galaxy 
properties typical of field to clusters takes place just at the characteristic 
densities of groups \citep{Lewis2002,Goto2003,Gomez2003,Marino2016}. 

In the following, we discuss our results considering substructures in the Dorado group,
from its {\it barycentre} to the periphery.

\subsection{The Dorado barycentre: the compact group SCG 0414-5559}
\label{Dorado-compact}

The barycentre of Dorado hosts  the compact group SGC~0414-5559 
identified by \citet{Iovino2002} (for the compact group definition 
see \citet{Hickson1982}). 
SGC~0414-5559  includes 4 members, namely NGC 1546 (member C),  
NGC 1549 (member B), NGC 1553 (member A) and IC 2058 (member D). The average
recession velocity of the four galaxies is 1259 km~s$^{-1}$ with a 
standard deviation  of  93 km~s$^{-1}$ (see Figure~\ref{fig_a1} and 
Table~\ref{Dorado_members}).

\begin{figure}[t]
\center
{\includegraphics[width=8.7cm]{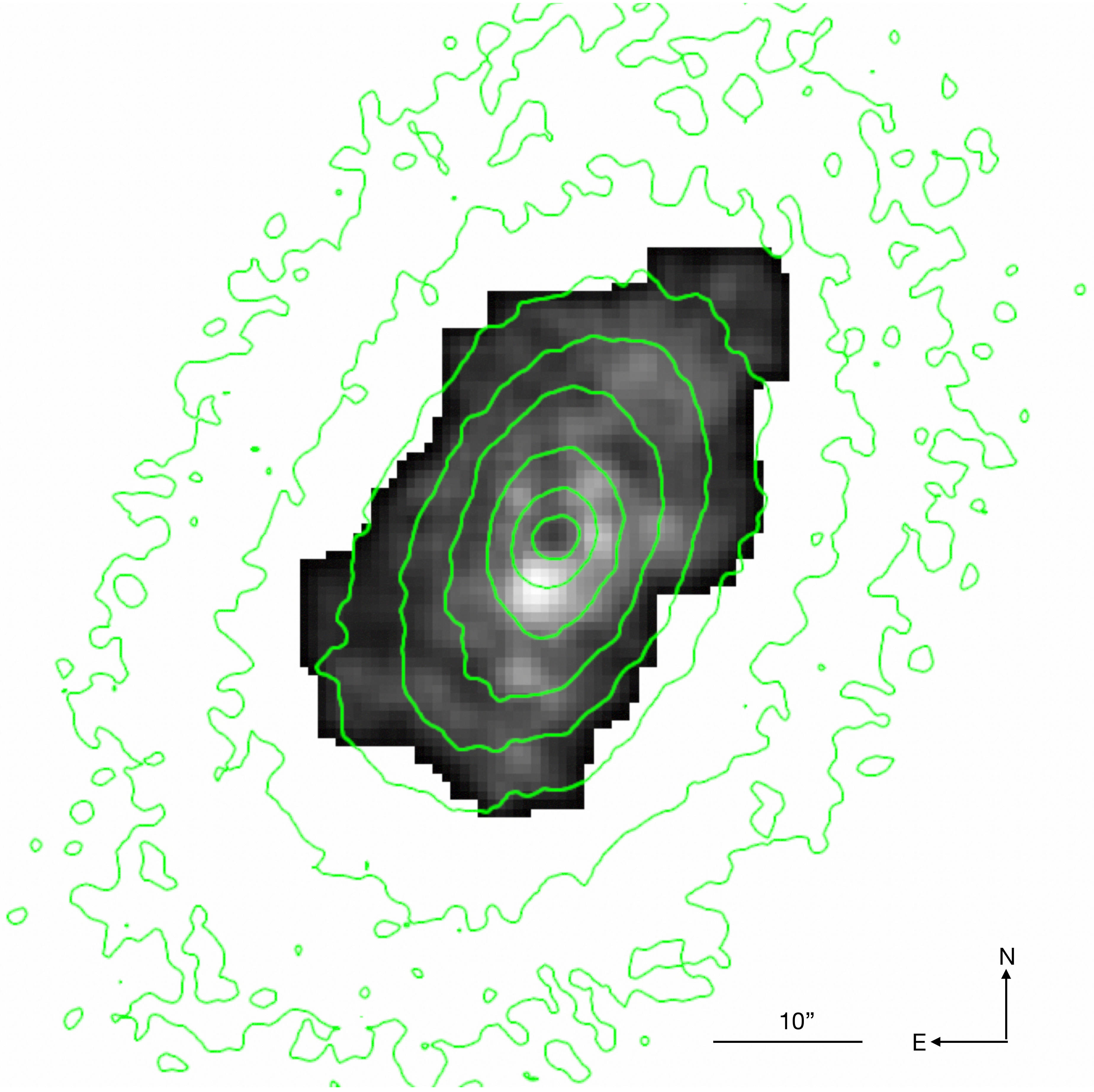}}
  \caption{NGC 1553. H$\alpha$ monochromatic map obtained from Fabry-Perot
 \citep{Rampazzo2003} overplotted to the present \HaN\ contours. 
 The \Ha\ emission is clumpy and extends for about 30\arcsec.
Note that the \Ha\ peak is displaced from the \HaN\ centre and corresponds
to the X-ray source  Chandra CXOU J041610.5-554646.8.
  \label{NGC1553_FP}}

\end{figure}

\begin{figure*}[t]
	\center
	{\includegraphics[width=13cm]{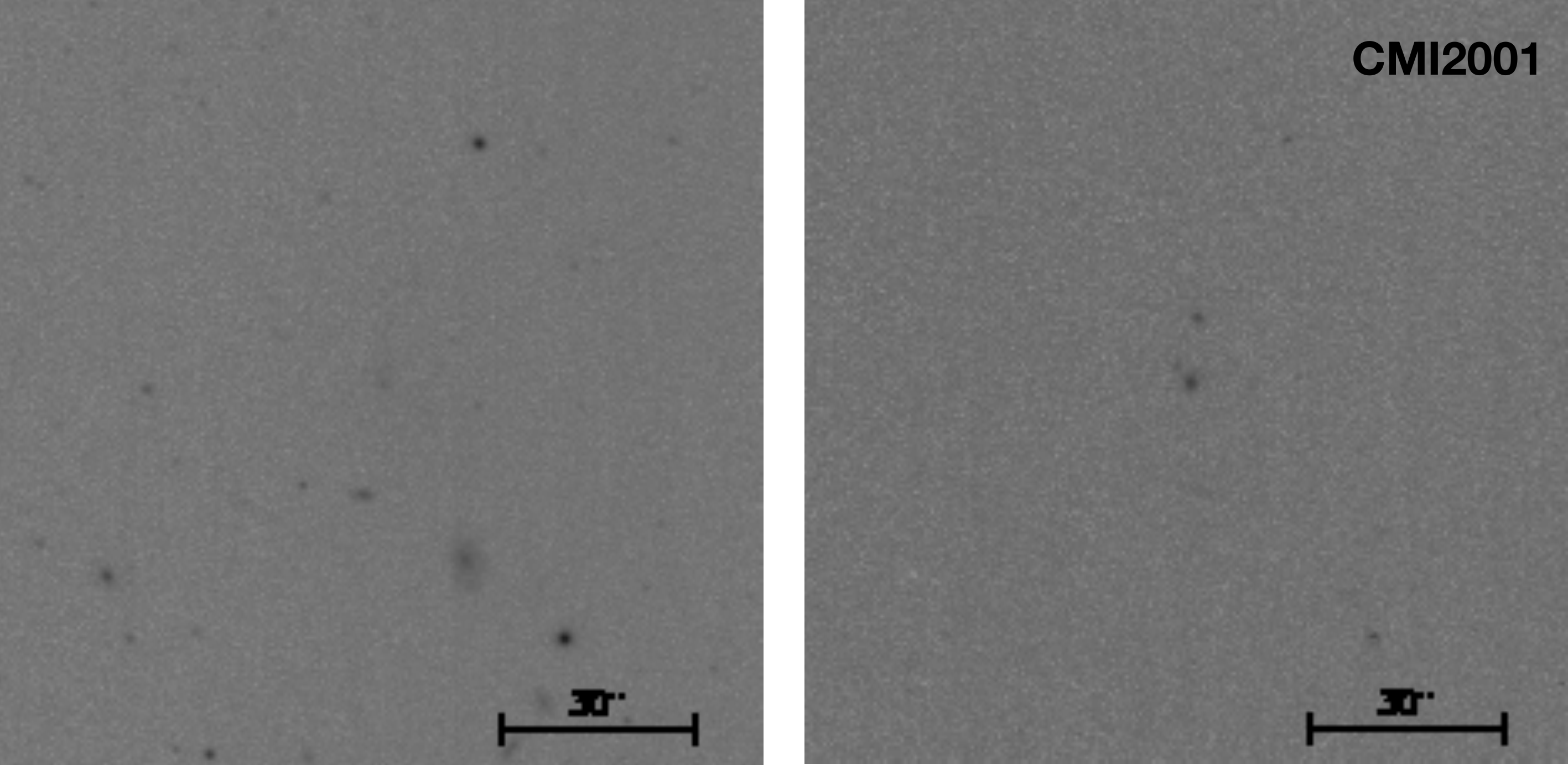}}
	\caption{[CMI2001]4136-01 region (RA=04 16 15.1 Dec=-55 41 51 J2000)  is
a region, North-East of NGC 1553, that has been detected by {\it GALEX}. A zoom
on this area  is shown in the continuum  (left panel) and in \HaN\ (right panel). 
The field of view is 2\arcmin$\times$2\arcmin. In the centre
of the field, two \HII\ regions are clearly visible in \HaN\ frame, 
while in the continuum objects barely detected.
		\label{CMI2001}}
\end{figure*}

\citet{Diaferio1994} proposed that compact groups form continually in a
single rich group during its collapse and virialization. This idea is based on
the observational evidence that compact groups are located within looser
structures or rich neighborhoods \citep[see e.g.][and references
therein]{Ribeiro1998}. According to \citet{Diaferio1994} N-body simulations, 
the survival time of these structures is about 1 Gyr and most of the 
member galaxies are not merger remnants. 

As sketched  in Section~\ref{Dorado-description}, galaxies in SCG 0414-5559
are, at odds, rich of merging and interaction signatures.
The nuclei of the two shell galaxies, NGC 1549 and NGC 1553 are LINERs 
 \citep[see e.g.][]{Annibali2010, Rampazzo2013}. 
Optical line strength indices found that NGC 1553 has a
young nucleus with a luminosity weighted age of 4.$\pm$0.7 Gyr
\citep{Annibali2007}. Mid infrared {\it Spitzer}-{\tt IRS} spectra of 
both NGC 1549 and NGC 1553 from \citet{Rampazzo2013} show that PAH are 
present in the nucleus with anomalous emission ratios. This kind of 
nuclei could host AGN as well as residuals of past star formation events 
\citep[see e.g.][]{Vega2010}. In the X-ray domain the emission of the nucleus of 
NGC 1553 is AGN-like \citep{Flohic2006}. {\tt GALEX} observations detected both
NGC 1553 \citep{Marino2011a} and NGC 1549  in NUV while a ring is barely 
visible in FUV  in NGC 1553.

Figure~\ref{NGC1549} and Figure~\ref{NGC1553} show that the
\HaN\ emission distribution, along the galaxy body, is following the stellar
continuum. \citet{Gavazzi2018}
	suggest that weak \Ha\ emitters, e.g. galaxies dominated by [N II], can harbor 
	AGN in their centre, not connected with the observed widespread SF. 
	Spectroscopic studies, present in the literature,
	illustrate and enrich the above finding.
\citet{Rampazzo2003} used Fabry-Perot observations
to study the \Ha\ kinematics in the central regions of NGC 1553. 
The high spectral resolution (R=9400 at \Ha, for S/N=3) of the instrument made it possible   
to isolate \Ha\ emission showing that gas and stars co-rotate. In Figure~\ref{NGC1553_FP}
we overplot the \Ha\ monochromatic map from the Fabry-Perot to our \HaN\ isophotal contours.
The map shows the clumpy and irregular \Ha\ distribution, whose peak is centred neither on
the continuum nor on the \HaN\ emission, but coincides with an X-ray discrete source
whose colours suggest it to be an absorbed AGN \citep{Blanton2001,Flohic2006}.
The X-ray discrete source,  Chandra CXOU J041610.5-554646.8, seems partly 
responsible for the very innermost \Ha\ emission of NGC 1553.

The outskirts of NGC 1553 are characterized by very extended shells \citep{Malin1983}. 
In the field observed we have searched for extragalactic \HII\ regions and dwarf
galaxies. [CMI2001]4136-0 is an object, well visible in {\it GALEX} FUV, projected on the 
NGC 1553 outskirts (Figure~\ref{CMI2001}). Although this object is classified as a galaxy 
in CDS, we we do not see any obvious galaxy in continuum frame. Rather, in \Ha\ we
see two \HII\ regions which should be associated to NGC 1553.

The region of IC~2058, South-East of NGC 1549/NGC 1553,
is  \HI\ rich \citep{Kilborn2009} with a mass of 
17.1$\pm$1.2 $\times$10$^{8}$ M$_{\odot}$. \citet{Pearson2016}
studied the  \HI\ distribution enclosing  IC 2058 and the dwarf galaxy 
PGC 75125. On the basis of the low outer vs. total \HI\ gas fraction
they suggest that most of the \HI\ gas, removed in the IC2058/PGC 75125 
galaxy-galaxy interaction,  is in an ionized state. 
Both IC~2058 and PGC 75125 show \HII\ complexes 
\citep[see also][for IC2058]{Rossa2003}.
The \HaN\ emission, crossing  PGC 75125 as a clumpy and warped lane, 
may suggest an {\it in fly} gas refueling 
\citep[see e.g.][and references therein]{Domingue2003} following 
the picture described by \citet{Pearson2016}.

NGC 1546 is located south west of NGC 1549 and NGC 1553. 
\citet{Kilborn2005} detected this galaxy in \HI\ with a  
mass of 24.1$\pm$1.4$\times$10$^{8}$ M$_{\odot}$. Its
 \HII\ regions have a flocculent appearance likely following
 the inner and outer rings described by \citet{Comeron2014}. 

Summarizing, the SF  is concentrated at the southern periphery of the
SCG 0414-5559 compact group in correspondence of  a rich concentration
of \HI, while the area covered by NGC 1549 and NGC 1553 is undetected 
down to the \citet{Kilborn2005} survey limit of $\approx$ 3.5$\times$10$^{8}$ M$_{\odot}$.
However, \Ha\ has been detected using Fabry-Perot in the central 30\arcsec\ by
\citet{Rampazzo2003} and in this paper as \HII\ regions in the NGC 1553 outskirts.

\subsection{The north west side of Dorado backbone: NGC 1533, IC 2038 and IC 2039}

The north west side of Dorado is a physical sub-structure in the group (see e.g. 
Figure~\ref{fig_a1} bottom panel).
\citet{Ryan2003} detected a wide plume of \HI\ extending 
from IC 2038 to the East towards NGC 1533. Several extragalactic \HII\ regions
have been detected in this \HI\ tail  by \citet[][their Figure 7]{Werk2010}. 

\begin{figure*}[t]
	\center
	{\includegraphics[width=13cm]{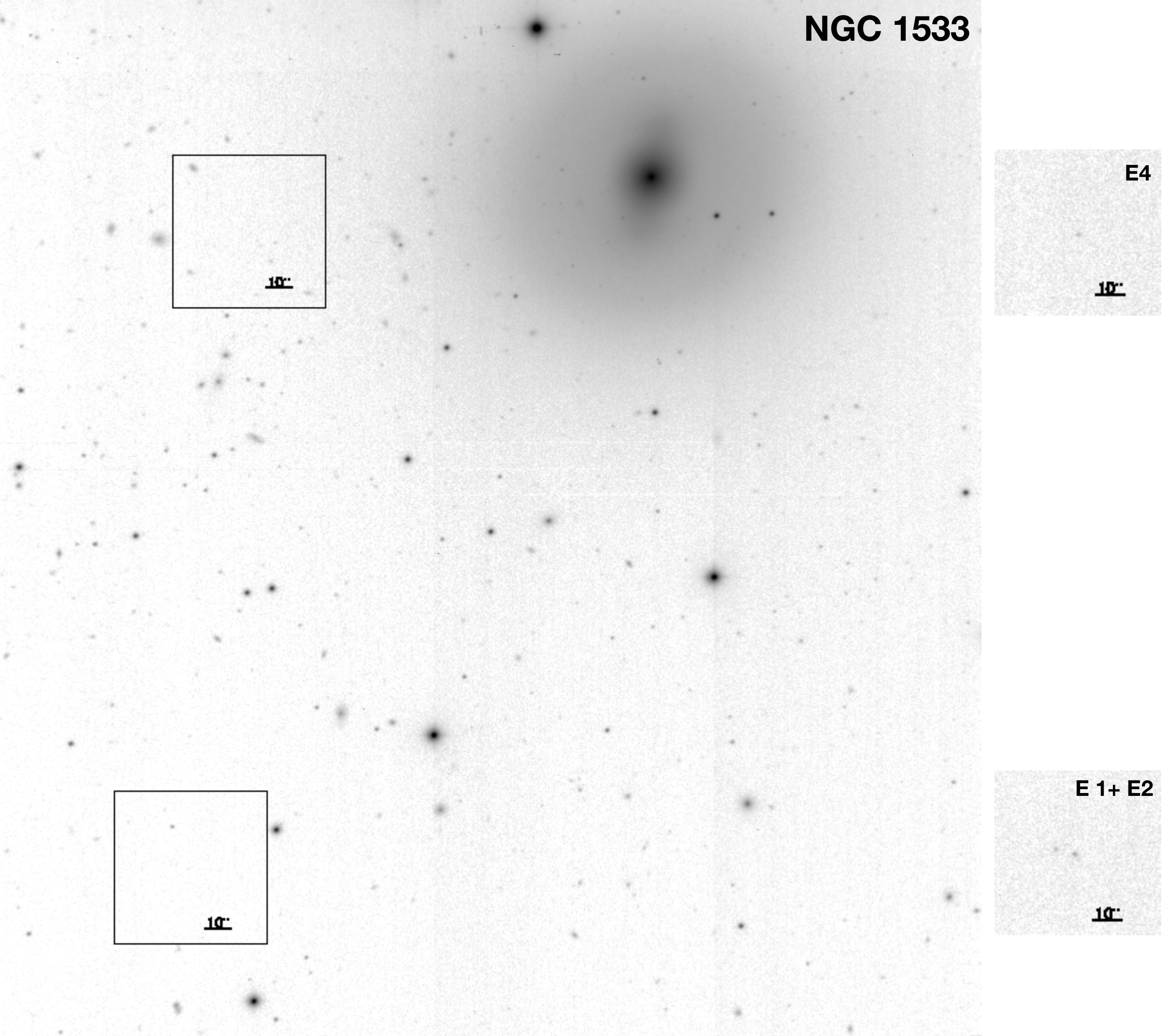}}
	\caption{({\it Left panel}) The region south-east of NGC 1533 shown in the continuum nearby 
      H$\alpha$. The two areas of  60\arcsec$\times$60\arcsec inserted in the panel, 
      mark the position of \HII\ regions,
revealed by our \HaN\ observations,   E1+ E2 (04 10 13.5 -56 11 36 J2000) and  
E4 (04 10 10.66 -56 07 27.99 J2000)  and shown in the two right panels. These \HII\ regions
have been firstly identified by \citet{Werk2010}.
		\label{NGC1533_E1}}
\end{figure*}

Figure~\ref{NGC1533}  shows that SF in NGC 1553 is not shut down:
\HII\ regions, indicated with A and B,  are outside the galaxy centre in a 
star forming ring shown by \citet{Marino2011b} in NUV and FUV.  
We have searched  for \HII\ regions detected by \citet{Werk2010}. 
The \HII\ regions shown in Figure~\ref{NGC1533_E1}, correspond
to the \citet{Werk2010} detection. 
However, the triplet deserves a further investigation 
with wide field, deep \HaN\ images \citep[see e.g.][]{Boselli2018}.

Our observations confirm that SF is detected not only in the spiral galaxy 
IC 2038 and in the centre of NGC 1533 but also along the \HI\ tails. IC~2038
SF is depressed if compared with objects of the same morphological type
(Figure~\ref{SFR_comparison}) likely due to the on-going stripping 
\citep[see e.g.][]{Werk2010,Cattapan2019}.

\subsection{The south-west edge of Dorado: NGC 1536 and NGC 1543}

NGC 1536 and NGC 1543 appear quite isolated in projection within the group,
although their heliocentric velocities (V$_{hel}$=1296 and V$_{hel}$=1148 \kms,
respectively) are very close to that of the central pair NGC 1549 and NGC 1553
(V$_{hel}$=1202 \kms). NGC 1536 structure is quite asymmetric.

Both NGC 1536 and NGC 1543 have been detected in \HaN, \HII\ regions
are revealed throughout the body of NGC 1536 but the galaxy has a very low
SFR (Figure~\ref{SFR_comparison}). \citet{Kilborn2005} noted that this
galaxy has ``5-10 times less \HI\ detected than expected''.

\HII\ regions are detected in the outer ring in 
NGC 1543.  In the  central 1.5\arcsec of NGC 1543, \citet{Parkash2019} 
measured a ratio [NII]/\Ha=2.004 confirming that, as in the case of NGC 1553, 
the dominant phenomenon at least in the nucleus, is not SF.
Indeed, the galaxy nucleus is classified as a LINER.
Although NGC 1543  is classified as a barred S0 (see Table~\ref{Dorado_members}),
\citet[][and references therein]{Murugeshan2019} included the galaxy 
in a sample of \HI-deficient late type galaxies (see also \citet{Kilborn2005}). 
Investigating the possible 
role of the angular momentum in the \HI\ depletion (log$_{\HI}$=8.75 M$_\odot$),  
they show that \HI\ is concentrated in the galaxy outskirts,
overlapping the stellar outer ring. The galaxy main body is devoid of \HI\ at the
limit of their survey. The \HII\ complexes we find in this galaxy 
develop at the south-east edge of the \HI\ and stellar ring. 
\citet{Murugeshan2019} suggested that the \HI\ hole in NGC 1543 is
regulated  by the galaxy  specific angular momentum rather than the
influence of the environment. 

\citet{Marino2011b} found FUV emission in the same area in which we
found \HII\ regions. \citet{Mazzei2019}  showed that  the global properties 
of this galaxy, both photometric and spectro-photometric, (SED, B absolute magnitude, 
morphology, velocity dispersion and rotation, X-ray luminosity of the hot gas 
and \HI\ gas amount)  can be accounted for by a minor merger event (mass ratio 5:1). 
The galaxy has a global age of 10.7 Gyr and suffered {\it rejuvenation} episodes 
in the last 2.3 Gyr. 

In these galaxies, the "\HI\ deficiency" seems dominant in regulating the
SF. 

\subsection{Towards east along the backbone: the pairs NGC 1566/NGC 1581 and
NGC 1596/NGC 1602 }

East of the SCG 0414-5559 central compact group, along
the Dorado backbone three objects are found in projection. NGC 1566,
NGC 1581 and the clump formed by NGC 1596 and NGC 1602. There is evidence
that not only the two close galaxies NGC 1596 and NGC 1602 form a physical pair but
also the very separated ones NGC 1566 and NGC 1581. Both regions are \HI\
rich \citep{Kilborn2005,Chung2006}.

\citet{Kendall2015} and  \citet{Oh2015} suggest that NGC 1566 and NGC 1581 
form a physical pair of galaxies, although their projected separation in terms
of galaxy size, R$_{proj}$/R$_{25}$, is 9.7 kpc. This pair, with a relative mass 
ratio of 0.05 and a radial velocity difference of 96 \kms, has
high tidal parameter (a measure of the strength of the tidal interaction)
according to \citet{Kendall2015}.   
The arm structures of NGC 1566 (Figure~\ref{NGC1566}) still show 
signature of an encounter  that might have occurred  with NGC 1581 
\citep{Oh2015}, while NGC 1581 does not show obvious signatures of interaction 
(see Figure~\ref{NGC1581}). However, NGC 1581, which shows \HII\ 
regions in the inner ring, could have acquired the gas from NGC 1566  
during its peri-galactic passage.
Several acquisition and gas-stripping scenarios have been discussed 
in the case of mixed, ETG+Spiral, pairs by \citet{Domingue2003}, while
mass transfer examples are in described in detail by \citet{Keel2004}.

The case of NGC 1596/NGC 1602 physical pair (Figure~\ref{NGC1596}) can be another
example of \HI\ gas transfer by a donor.
The area in which the pair is found is very rich of neutral hydrogen \citep{Chung2006}
which extends from NGC 1602, the donor, to NGC 1596, the receiver. In NGC 1596,
the acquired gas counter-rotate with respect to stars. However, we do not detect 
\HII\ regions in NGC 1596 but only an extended \HaN\ emission. 
\citet{Chung2006} reported that the ionized gas in NGC 1596 produces
 mostly [OIII]$\lambda$5007 emission in the central part \citep[see also][]{Bureau2006}. 
 No mention about \Ha\ is made by \citet{Bureau2006}, although one channel 
 (of the dual-channel spectrograph) is centred on the \Ha\ wavelength.
The [OIII] emission is concentrated in the central regions 
(few arcsec, see Figure~1 in \citet{Bureau2006}).

\section{Summary and conclusions}
\label{Conclusions}

Dorado is a nearby, rich group extending for about 10 square degrees in
the Southern Hemisphere. We observed, in narrow-band \HaN\ imaging 
the 14 galaxies, both early and late types, that form 
the group backbone.  We obtained their \Ha\ luminosity and estimated 
their SFR. 

We obtained the following results:
\begin{itemize}

\item{All members have been detected in \HaN, irrespective of their
morphological type.}

\item{\HII\ regions are clearly visible in half of ETGs, namely NGC 1533, 
NGC 1543, NGC 1546, NGC 1581.  \HII\ regions are found in rings and/or filaments. 
In NGC 1549, NGC 1553, NGC 1596 the \HaN\ emission does not show the clumpy structures 
of \HII\ regions, but a rather smooth structure. However, \Ha\ is likely present 
in the galaxy central regions, as suggested by the high resolution Fabry-Perot 
observations of NGC 1553 \citep{Rampazzo2003}}.

\item{We detect \HII\ regions in the galaxy outskirts of
NGC 1533 \citep[][and reference therein]{Werk2010}  and in 
NGC 1553, detected also by {\it GALEX} as FUV-emitting regions. }

\item{\citet{Gavazzi2018}, for the 
	ATLAS$^{3D}$ sample detected 55/147 ETGs  i.e 37\%$^{+6}_{-11}$ of their sample
	(errors are calculated using 1~$\sigma$ low and upper limits from Poisson statistics
	following \citep{Gehrels1986}). Considering errors,  
\citet{Gavazzi2018} detected in \HaN\ less than half of their galaxies, likely because
of the significant presence of Virgo members which include  
passively evolving ETGs \citep[see e.g.][]{Bressan2006}. We conclude that 
ETGs in Dorado are leaving an active phase and their SF is not extinguished yet.}

\item{In the LTG, NGC 1536 the \HII\ regions are irregularly distributed 
following the morphological perturbation of the underlying galaxy. This is partially
seen also in IC 2038.}

\item{The EW of \HaN\ emission, a measure of the specific star formation rate, 
	 is increasing with the morphological type, although most of the
LTGs in our sample are below the mean and median values corresponding to their 
morphological types \citep{James2004}. The EW of ETGs extends the
trend towards low morphological type values, in the $-5\leq $Type$\leq$0 range.} 

\item{The SFR of the Dorado spiral members  is in the range of general surveys of 
LTG  \citep{James2004}, but rarely above the median  of their morphological type. 
Three galaxies, namely NGC 1536, PGC 75125 and IC 2058, have a SFR well below 
the median for their morphological classes.

The dominant mechanisms in action in this evolutionary phase of
the Dorado group  are gas stripping and gas  exchange via in flight re-fuelling between 
galaxies. Some galaxies are already "\HI\ deficient", as in cluster counterparts.}

\end{itemize}
 
\bigskip
Summarizing, the present \HaN\ observations  show the Dorado backbone 
to be a strongly evolving environment. \Ha\ is a short time scale (10$^{7}$ years)
indicator in terms of evolution. The gas reservoirs, still present in LTGs and
their surrounding environment,  sometimes trigger a residual
activity  in ETGs. 

The donor-receiver mechanism, via galaxy-galaxy
interaction, appears important at the edge of Dorado, considering
the pairs (NGC 1566/ NGC 1581, NGC 1596/ NGC 1602) and triplets 
(NGC 1533, IC 2038, and IC 2039).  Gas stripping is another
mechanisms in action:  NGC 1533 stripped IC 2038, and the same 
might have happened in the past between 
NGC1566 and NGC1581 and between NGC1596 and NGC1602.
The consequence of this stripping could be linked to the 
low SFR of the spiral IC 2038 (Figure~\ref{SFR_comparison}), 
to the presence of \HII\ regions in the inner ring of NGC1581 and 
to the \HI\ plume connecting  NGC1602 to NGC1596 \citep{Chung2006}.

The barycentre of the group, the compact group SCG 0414-5559, 
appears more evolved. In the two shell galaxies NGC 1549 and NGC 1553 the \HaN\ 
emission is dominated by [NII], although \Ha\ has been found in the NGC 1553 central region
via Fabry-Perot high resolution observations \citep{Rampazzo2003}. SF is still found 
in the outskirts
of this compact group in IC 2058, in its dwarf physical companion PGC 75125, both
enclosed in an \HI\ cloud, and in NGC 1546. 

The analysis of the SFR in Dorado will continue in the Far Ultraviolet.
The present sample has been already observed with {\tt UVIT} on {\tt Astrosat}
\citep{Tandon2017} in the FUV band. Their  UV SFR and its connection with 
\Ha\ emission
will be the subject of a forthcoming paper.

\begin{acknowledgements}
We wish to thank the unknown referee for very constructive suggestions.
R.R. and P.M  acknowledge the partial support of the  INAF PRIN-SKA
2017 program 1.05.01.88.04. MS acknowledges financial support from the VST 
project (P.I. P. Schipani). 
E.C. acknowledges support from ANID project Basal AFB-170002.
This paper includes data gathered with the du Pont and Swope  
Telescopes located at Las Campanas Observatory, Chile. We acknowledge the usage of the {\tt
HyperLeda} database ({\tt http://leda.univ-lyon1.fr}). {\tt IRAF} is distributed 
by the National Optical Astronomy Observatories, which is operated by the 
Association of Universities for Research in Astronomy, Inc. (AURA) under 
cooperative agreement with the National Science Foundation. 
\end{acknowledgements}

\bibliographystyle{aa} 
\bibliography{aa-Dorado.bib} 


\begin{appendix} 

\section{Dorado members}
\label{Dorado members}

 The Dorado group is indicated as PGC1 14765, i.e. with the PGC
 (Principal Galaxy Catalogue) number of the main galaxy  NGC 1553 in
 \citet{Kourkchi2017}  and counts 31 members. They are indicated with
 the number  N, reported in col.1 of Table~\ref{Dorado_members}. From
 this list we removed 2MASXJ04105983-5628496, which corresponds to NGC 1536
  that is already present in the list, reducing the number of members to 30.
 In Table~\ref{Dorado_members} we add, to the re-defined 
 \citet{Kourkchi2017} list, PGC 75125 from the \citet{Firth2006} Dorado sample.
 PGC 75125 is in the same frame of IC 2058 (see  Table~\ref{table-Ha-observations} 
 and Figure~\ref{Dorado-group}) and is a physical companion of this galaxy.

From the \citet{Kourkchi2017} paper, the table reports:

col. 1 member number;

col. 2 Right Ascension;

col. 3 Declination;

col. 4 PGC number;

col. 5 Galaxy name;

col.6 Morphological Type;

col. 7 the total apparent B-band magnitude;

col. 8 the total apparent K$_s$-band magnitude;

col. 9 the logarithm of the total apparent K$_s$-band magnitude in solar luminosity;

col. 10 the heliocentric recession velocity;

col. 11 the galaxy distance in Mpc and the percentage error.

\begin{figure}
{\includegraphics[width=8.9cm]{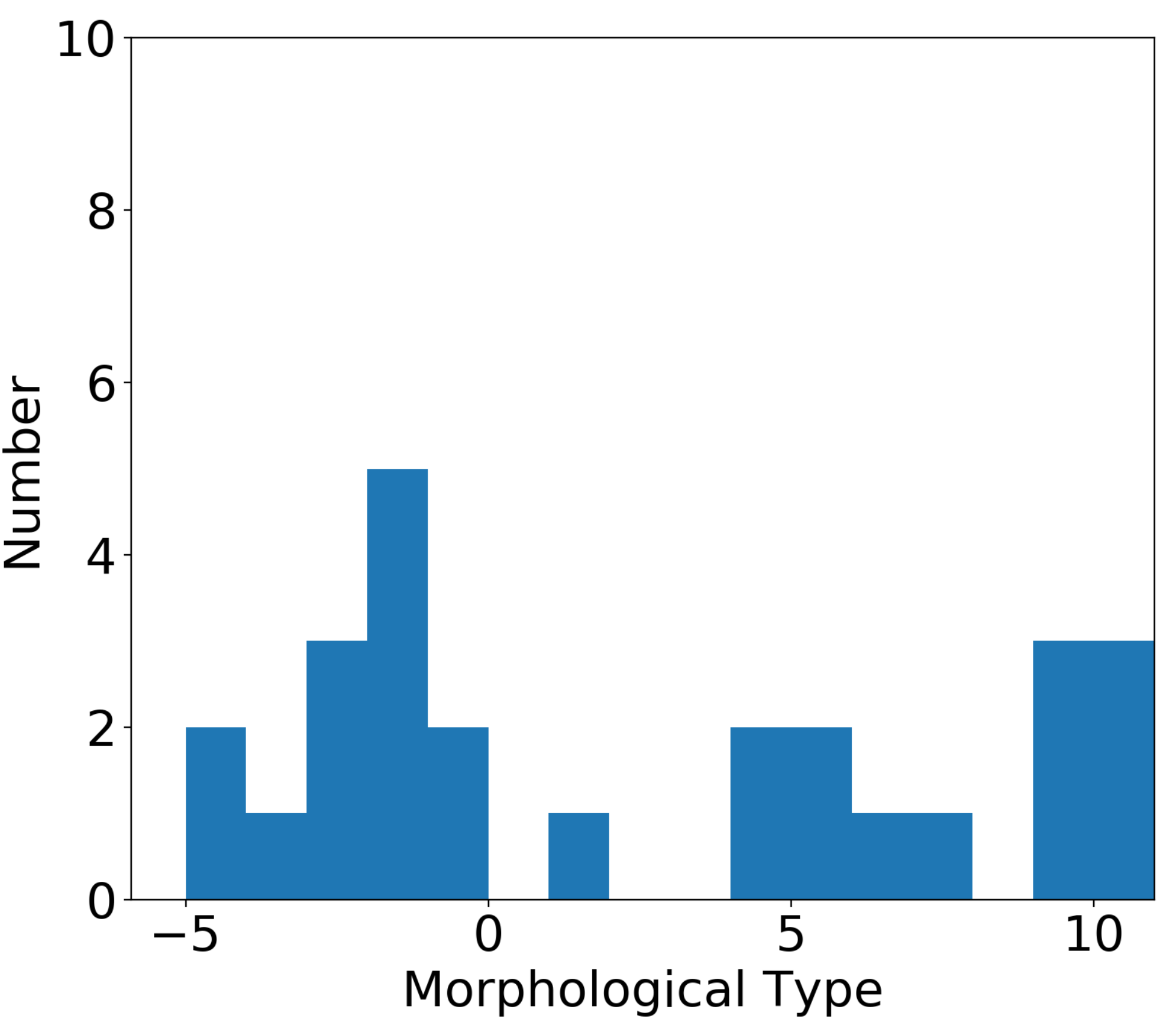}
\includegraphics[width=8.9cm]{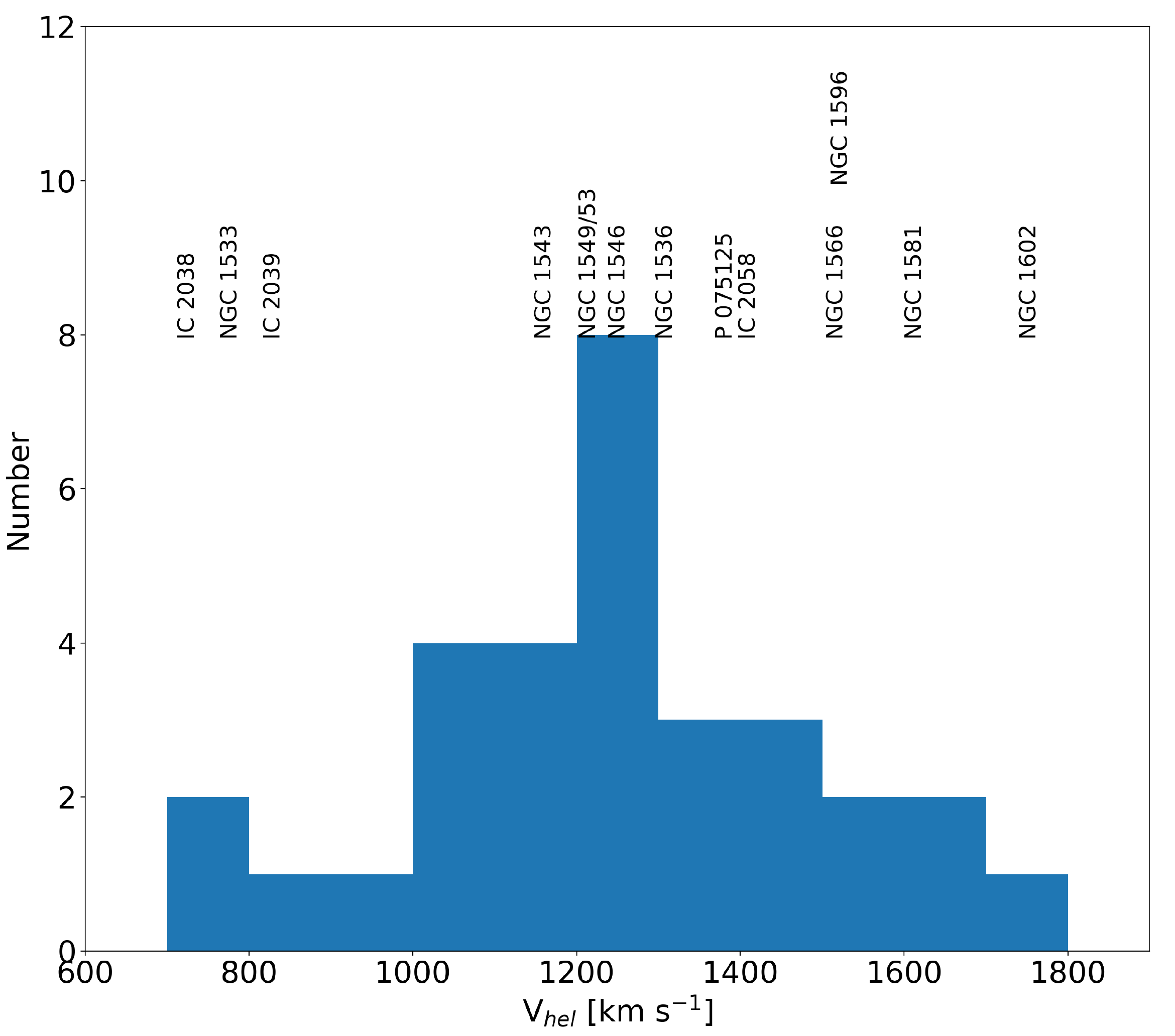}}
\caption{From top to bottom, morphological type (only 26/31 objects 
in Table~\ref{Dorado_members} have a morphological classification) and  heliocentric velocity
distribution of Dorado members in Table~\ref{Dorado_members}. Members included 
in this study are indicated in correspondence of their heliocentric velocity.
\label{fig_a1}}
\end{figure}

The distribution of the morphological type and of the heliocentric velocity
are shown in Figure~\ref{fig_a1}.

\begin{sidewaystable*}
\caption{Dorado members.}\label{Dorado_members}
\centering
\begin{tabular}{lllrlcccccc} 
\hline\hline   
N   &   RA        &     Dec      &   PGC  & Galaxy                 &      Morpho.  &  B     &    K$_s$   &  log K$_s$          & V$_{hel}$             &  Distance   \\
      &   [J2000]    &  [J2000]  &             &    Identification   &      Type    & [mag]& [mag] & [L$_\odot$] & km~s$^{-1}$ & [Mpc] (error)\\
\hline
26 &	04 01 15.240 &	-53 29 23.28 &	429411 & PGC429411	            &  10.0   &	16.18 &	14.26 &	7.87 &	1135 & 	\\	
 8 &	04 04 02.712 &	-54 06 00.00 &	14397 &	NGC1515	                    &  4.0    &	10.87 &	7.93 & 10.40 &	1175 &17.54(14\%)     \\	
15 &	04 07 03.792 &	-55 19 27.48 &	14481 &	IC2032	                    &  9.9     &	14.08 &	10.86 &	 9.23 &	1066 & 	  	  \\	
19 &	04 08 53.760 &	-55 59 22.20 &  {\bf 14553} & {\bf IC2038}	                    &	 7.0 &	14.31 &  12.00 & 8.78  & 712 & 	         \\	
22 &	04 09 02.376 &	-56 00 42.12 &	{\bf 14560} &	{\bf IC2039}	                    &	-3.1 &	14.92 &	12.41 &	8.61 &	817 & 	         \\	
  7 &	04 09 51.840 &	-56 07 06.60 &	{\bf 14582} &	{\bf NGC1533}	                    &	-2.5 &	11.71 &	7.74 &  10.48 &	764 & 20.89(11\%)      \\	
 11 &04 11 00.528 &  -56 29 05.64 & {\bf 14620}   & {\bf NGC 1536}                       & 5.0  &       13.19   & 9.84       & 9.64     &     1296  &  \\
 6 &	04 12 43.200 &	-57 44 15.72 &	{\bf 14659} &	{\bf NGC1543}	                    &  -2.0 &	11.36 &	7.55 &	10.55       &	1148 &18.71(11\%)      \\	
 23 &04 14 20.304 &	-58 12 26.2 &  381152 & PGC381152                     &        &   16.61 &	12.43 &	 8.60  &1233    &	  	  \\	
 10 &04 14 36.456 &	-56 03 39.24 &	{\bf 14723} &	{\bf NGC1546}	                  &	-0.4 & 11.95 &	8.16  &	10.31 &	1238 &	  	  \\	
 21 &04 14 40.872 &	-58 07 55.20 &	75108 &	PGC075108/6DF         &  	-2.0 &	16.36 &	12.06 &	8.75 &	1242 &	 	  	   \\	
 2 &	04 15 45.144 &	-55 35 32.28 &	{\bf 14757} &	{\bf NGC1549}	                    &  	-4.3 &	10.61 &	6.88 &	10.82 &	1202 &17.38(10\%)	  \\		
 1 &	04 16 10.488 &	-55 46 48.00 &	{\bf 14765} &	{\bf NGC1553}	                    &  	-2.3 &	10.20 &	6.34 &	11.04 &	1201 &15.63(11\%)	  \\		
16 &	04 17 54.336 &	-55 55 58.08 &	{\bf 14824} &	{\bf IC2058}	                    &  	 6.5 &	12.12 &	10.98 &	 9.18 &	1397 &	 	  	   \\	
18 &	04 18 59.472 &	-58 15 27.36 &	14850 &	ESO118-019	            &  	-0.3 &	14.83 &	11.53 &	 8.96 &	1239 &	 	  	  \\		
 3 &	04 20 00.384 &	-54 56 16.08 &	{\bf 14897} &	{\bf NGC1566}	                    &  	 4.0         &	 9.98 &	7.03 &	10.76 &	1504 &		 	  	  \\			
 5 &	04 21 58.824 &	-56 58 29.28 &	14965 &	NGC1574	                     &	-2.9 &	11.23 &  7.23 &	10.68 &	1041 &	   19.32(11\%)	  \\		
25 &	04 22 51.720 &	-56 13 35.04 &   75146 &APMBGC157+016+068    &     10.0 &	15.45 & 13.82 &	 8.05 &	1345 &		 	  	  \\		
 12 &	04 24 44.952 &	-54 56 31.20 &	{\bf 15055} &	{\bf NGC1581}	                    &	-2.9 &	13.51 &	9.94 &	 9.60 &	1600 &		 	        \\	
29 &	04 25 20.208 &	-56 49 23.88 &   75152 & PGC075152	                    &  	      &       18.64 & 14.59   &	 7.74 &	1169 &	  	  	  \\		
30 &	04 25 25.632 &	-56 44 25.44 & 3315626 & PGC3315626	            &  	 &  19.13 & 15.03 &	 7.56 &	1499 &	 	  	  \\		
31 &	04 26 39.168 &	-56 53 44.16 & 3315643 & PGC3315643	            &  	 &  19.28 & 15.28 &	 7.47 &	1049 &	 	  	  \\		
27 &	04 27 13.824 &	-57 25 41.88 &  390135 & PGC390135	            &  	10.0 &	16.65 & 14.31 &	 7.85 &	1215 &	 	 	  	  \\	
17 &	04 27 32.592 &	-54 11 48.12 &	15149 &	ESO157-030	            &  	-4.3 &	14.50 & 11.40 &	 9.01 &	1471 &		 	  	  \\		
9  &	04 27 38.088 &	-55 01 39.72 &	{\bf 15153} &	{\bf NGC1596}	                    &  	-2.0 &	11.94 &	8.10 &	10.33 &	1510 &   15.35(13\%)  \\		
28 &	04 27 53.640 &	-56 19 39.36 &  401051 & PGC401051	            &    	 &  17.24 &  14.48 & 7.78 &	1469 &	 	  	  \\		
20 &	04 27 54.984 &	-55 03 28.08 &	{\bf 15168} &	{\bf NGC1602}	                    &   9.5 &	12.86 &	12.02 &	 8.77 &	1740 &	 	  	  \\	
14 &	04 31 24.240 &	-54 25 00.84 &	15388 &	IC2085	                    &         -1.2 &         13.89 &	10.55 & 9.35     &	 982 &	  20.99(18\%)   \\		
 4 &	04 31 39.696 &	-54 36 07.20 &	15405 &	NGC1617	                    &      1.1   &	10.85 &	 7.16 &	10.71 &	1063 &	 	 	  	  \\	
24 &	04 37 18.024 &	-55 55 23.52 &	15661 &	ESO157-044	            &       9.9   &	14.45 &	12.59 &	 8.54 &	1611 &		 	        \\			
 *  & 04 18 07.100 &  -55 55 50.00 &    {\bf  75125}  & {\bf J04180709-5555503}          &      5.0             &                 &             &             &     1369  &                                        \\	
\hline
\end{tabular}
\tablefoot{Galaxies of \citet{Kourkchi2017} list observed are emphasised in bold in 
		columns  4 and 5. To the candidate list of \citet{Kourkchi2017} we add PGC 75125 (indicated
	with an asterisk) from \citet{Firth2006} we observed and analyzed in this work.}
\end{sidewaystable*}

\section{Comparison with the literature}
\label{comparison_literature}

Only one galaxy, NGC 1566, in our sample has been previously
observed in the same band. \citet{Hoopes2001} and \citet{Kennicutt2009} performed 
respectively \Ha\ and \HaN\ imaging of the entire galaxy. 
\citet{Hoopes2001} observed at the CTIO 0.9m telescope using \Ha\ 6602/20 
filter and an R-band filter to remove the stellar continuum. The 6602/20 filter,
centred at $\lambda$=6596 \AA\ has a FWHM=18\AA\ and a transmission of 70\%
at the central wavelength.  As a consequence, the [NII] emission does not
contaminate the measure. The paper reports the \Ha\ luminosity
 $L_{H\alpha}$=31.65$\pm$0.6$\times$10$^{33}$ W 
 (adopting a distance of 17.49 Mpc). This is equivalent to a flux of 
880$\times$10$^{-14}$ erg~s$^{-1}$~cm$^{-2}$.

In their Table~1, \citet{Kennicutt2009} reported a measure of 
log(\HaN)=-10.9$\pm$0.1 erg~s$^{-1}$~cm$^{-2}$ 
corresponding to a \HaN\ flux of 1318 $\times$10$^{-14}$erg~s$^{-1}$~cm$^{-2}$. 

\citet{Kennicutt2009} adopted the  [NII]/\Ha=0.62$\pm$0.06 ratio, 
to correct for the [NII] emission, 
obtained by \citet{Hawley1980} from the study of 3 \HII\ regions.
\citet{Roy1986} performed long-slit scanning spectroscopy at low resolution 
($\Delta\lambda$=0.7~\AA) covering a rectangular region of 120"$\times$16" 
positioned on the NW spiral arm of NGC 1566. 
They measured the lines fluxes in 5 \HII\ regions. From their work we derived
the average ratio [NII]/\Ha=0.38 significantly reduced with respect to \citet{Hawley1980}.

Assuming as the net \Ha\ flux the value of \citet{Hoopes2001}, 
the \HaN\ flux is 
1425.6$\pm$50.2  $\times$10$^{-14}$erg~s$^{-1}$~cm$^{-2}$ and
1214.4$\pm$33.4   $\times$10$^{-14}$erg~s$^{-1}$~cm$^{-2}$ adopting
the \citet{Hawley1980} and \citet{Roy1986}  [NII]/\Ha ratios, respectively. 

We conclude that our value 1246.7$\pm$125.1 $\times$ 10$^{-14} $erg~s$^{-1}$~cm$^{-2}$
(Table~\ref{HA-measures}) agrees, within
errors, with both \citet{Hoopes2001} \Ha\ estimate, 
once adopted the \citet{Roy1986} [NII]/\Ha ratio, and 
\citet{Kennicutt2009}.

\section{Images in the continuum and in the \HaN\ filter}

We present here the images of the galaxies observed in \HaN. 
For galaxies larger than the du Pont field of view (8\farcm85$\times$8\farcm85)
we obtained a mosaic of images. Left and right panels show
continuum and  H$\alpha$ images, respectively. Stellar residuals, sometimes 
present in the H$\alpha$ images due to bright stars, have been removed.

\begin{figure*}[t]
\center
{\includegraphics[width=15cm]{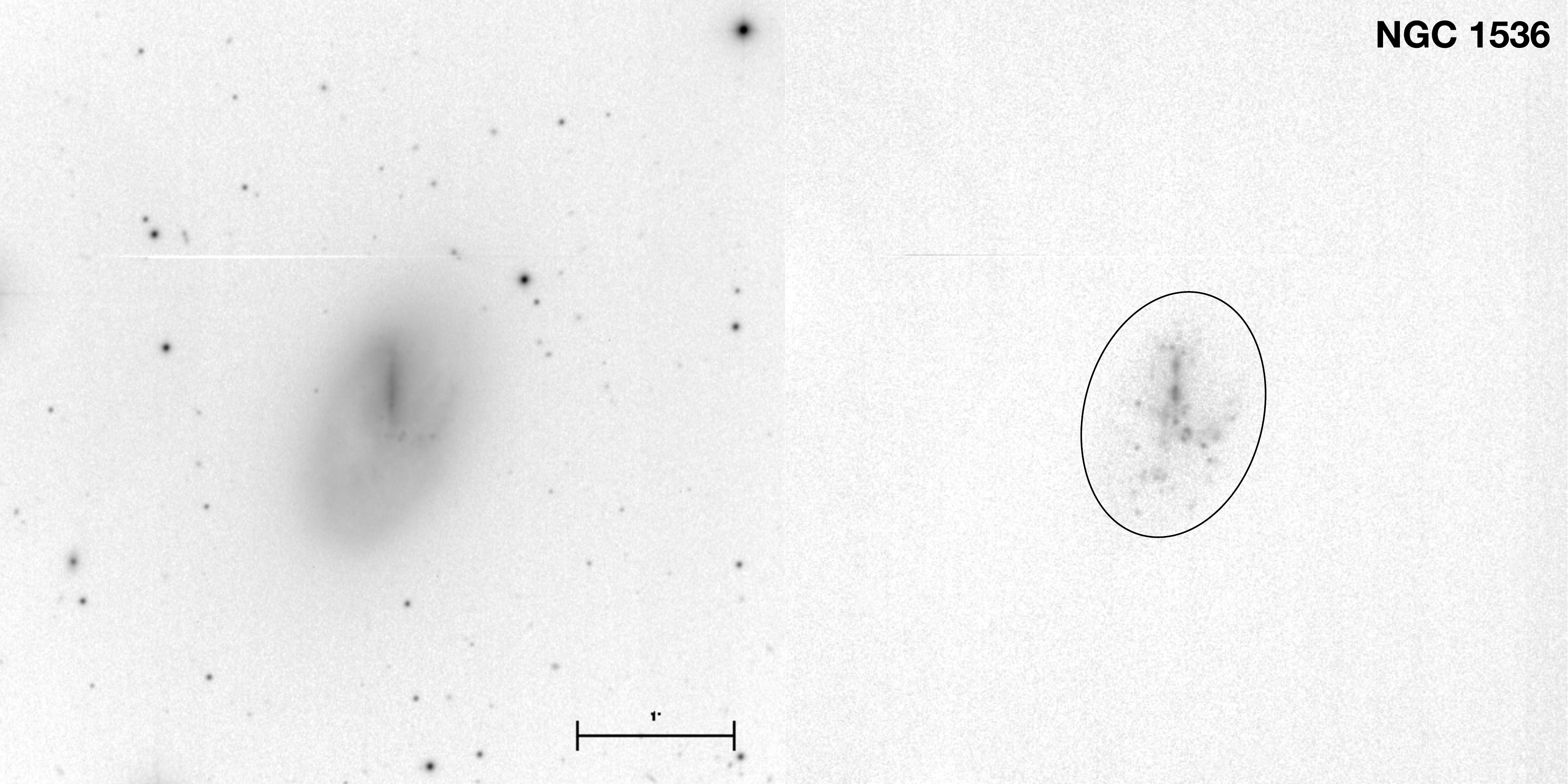}}
{\includegraphics[width=15cm]{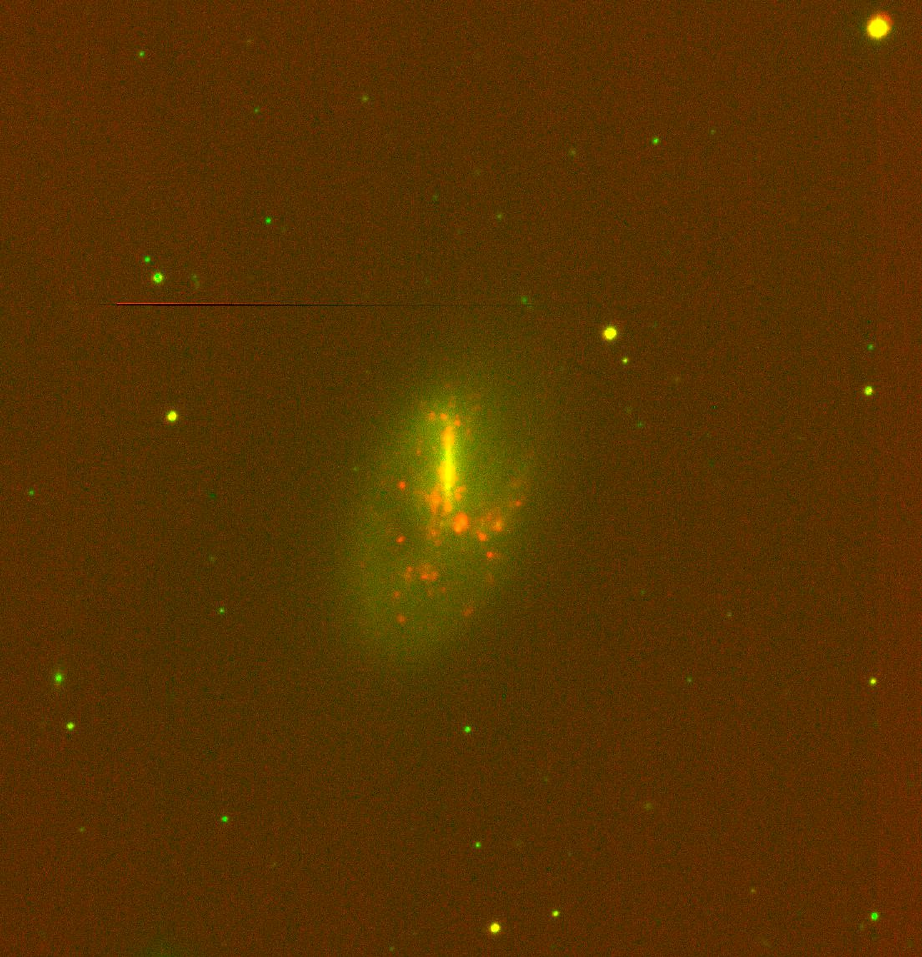}}
\caption{As in Figure~\ref{IC2038} for NGC 1536. 
The image size is  5\arcmin$\times$5\arcmin.} 
\label{NGC1536}
\end{figure*}

\begin{figure*}[t]
	\center
{\includegraphics[width=15cm]{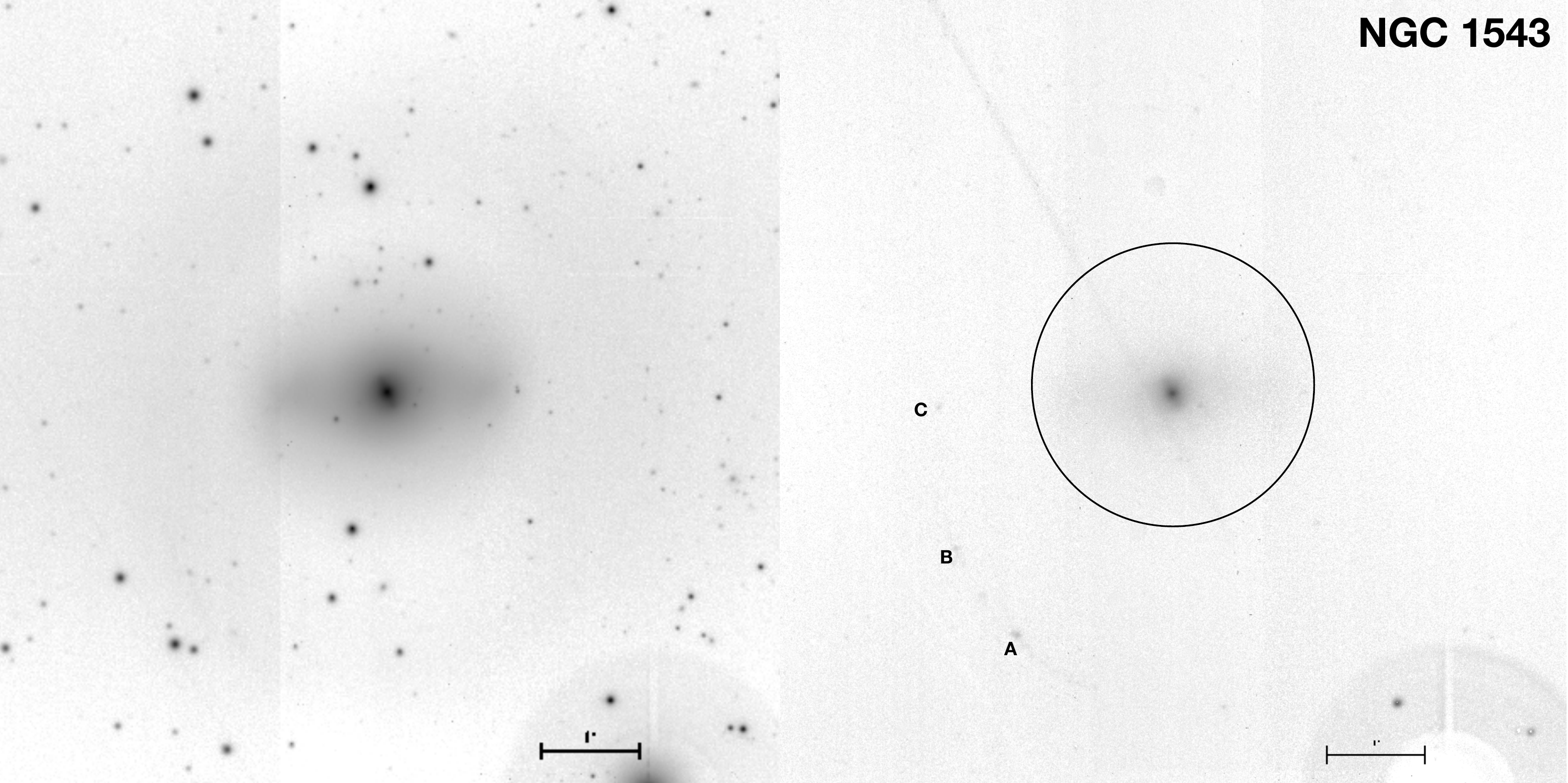}}
{\includegraphics[width=15cm]{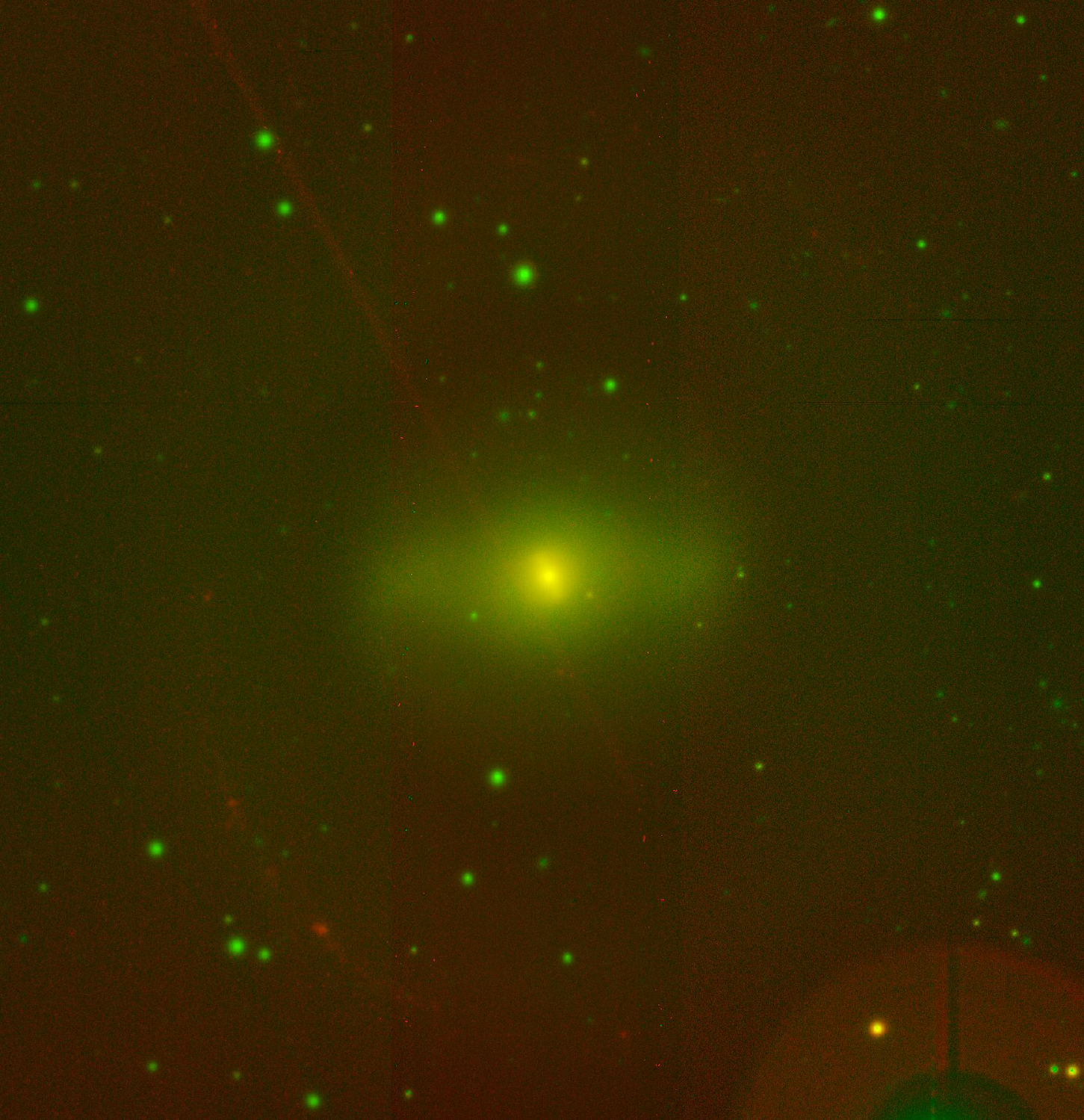}}
\caption{As in Figure~\ref{IC2038} for NGC 1543.
Labels A, B and C (top right panel) indicate emission areas along 
the outer ring of the galaxy. The total flux in the
in the ring is 3.72$\pm$0.37$\times$10$^{-14}$ erg s$^{-1}$ cm$^{-2}$ .
The image size is  8\arcmin$\times$8\arcmin.} 
\label{NGC1543}
\end{figure*}
%
\begin{figure*}[t]
\center
{\includegraphics[width=15cm]{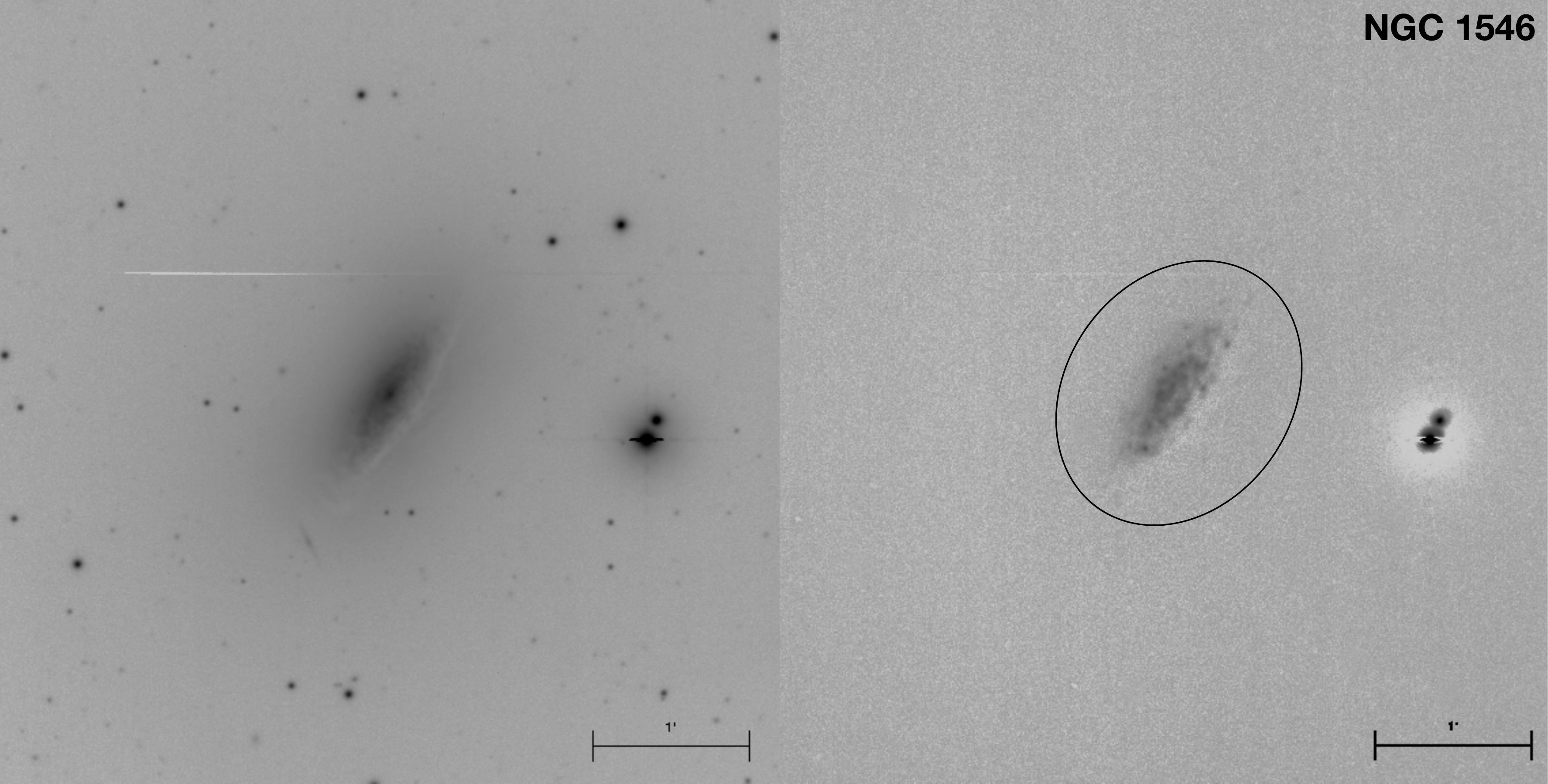}}
{\includegraphics[width=15cm]{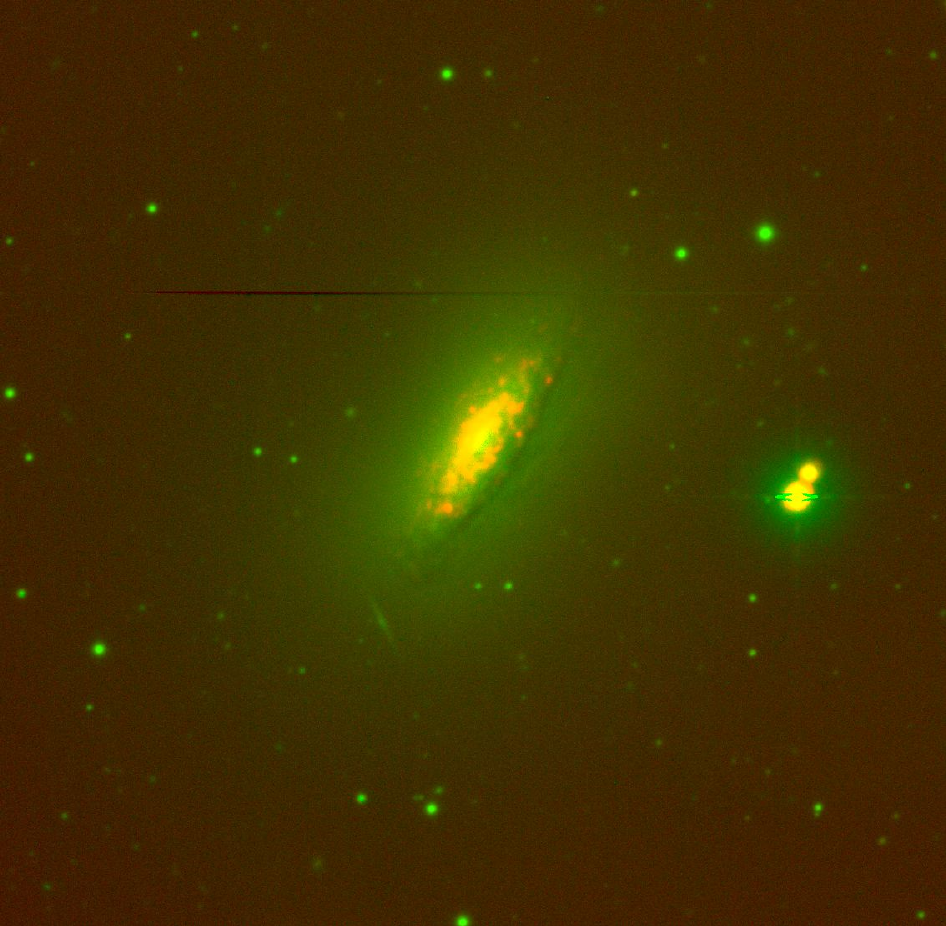}}
\caption{As in Figure~\ref{IC2038} for  NGC 1546. 
The image size is  5\arcmin$\times$5\arcmin. } 
\label{NGC1546}
\end{figure*}
%
\begin{figure*}[t]
\center
{\includegraphics[width=15cm]{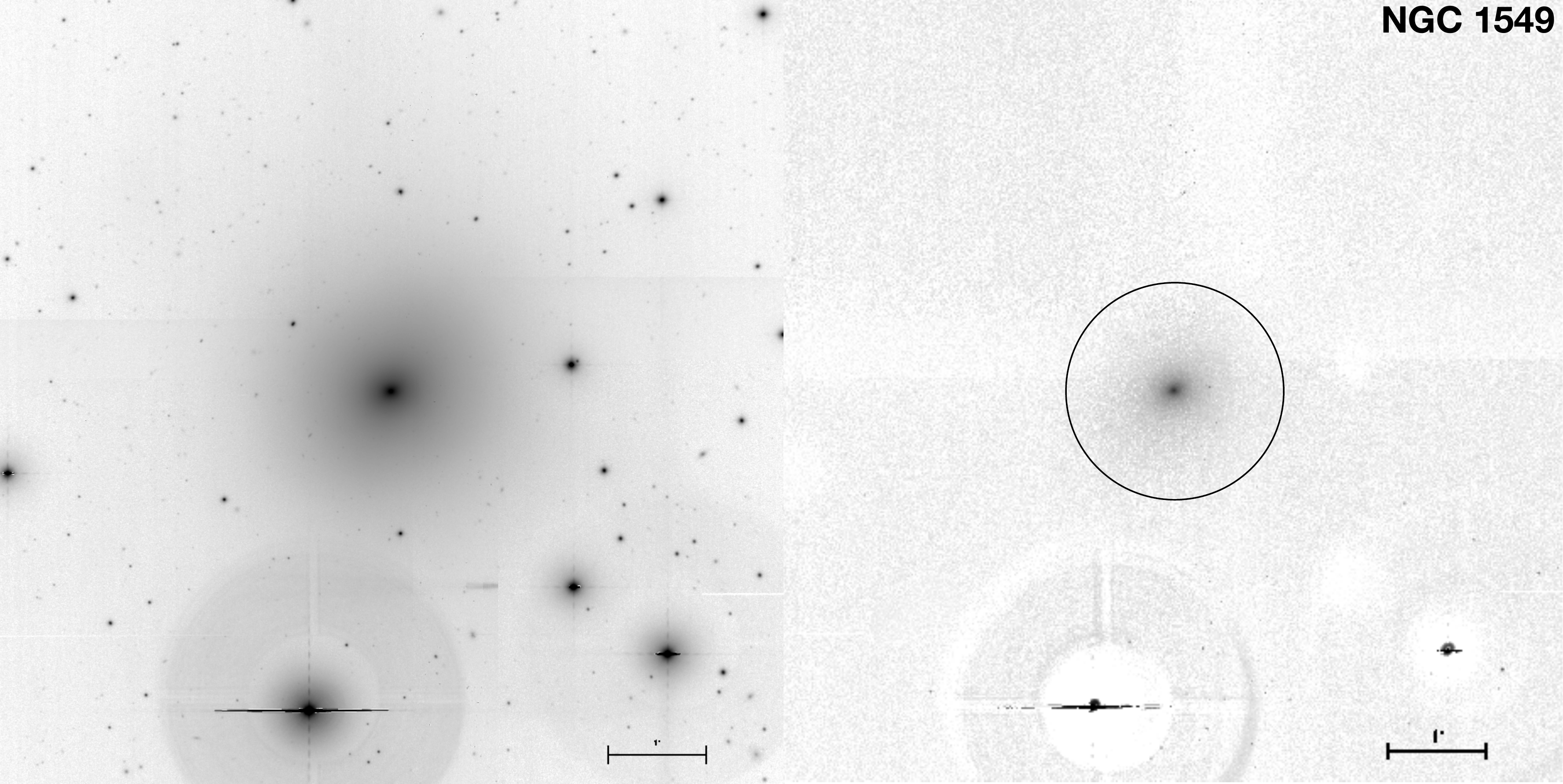}}
{\includegraphics[width=15cm]{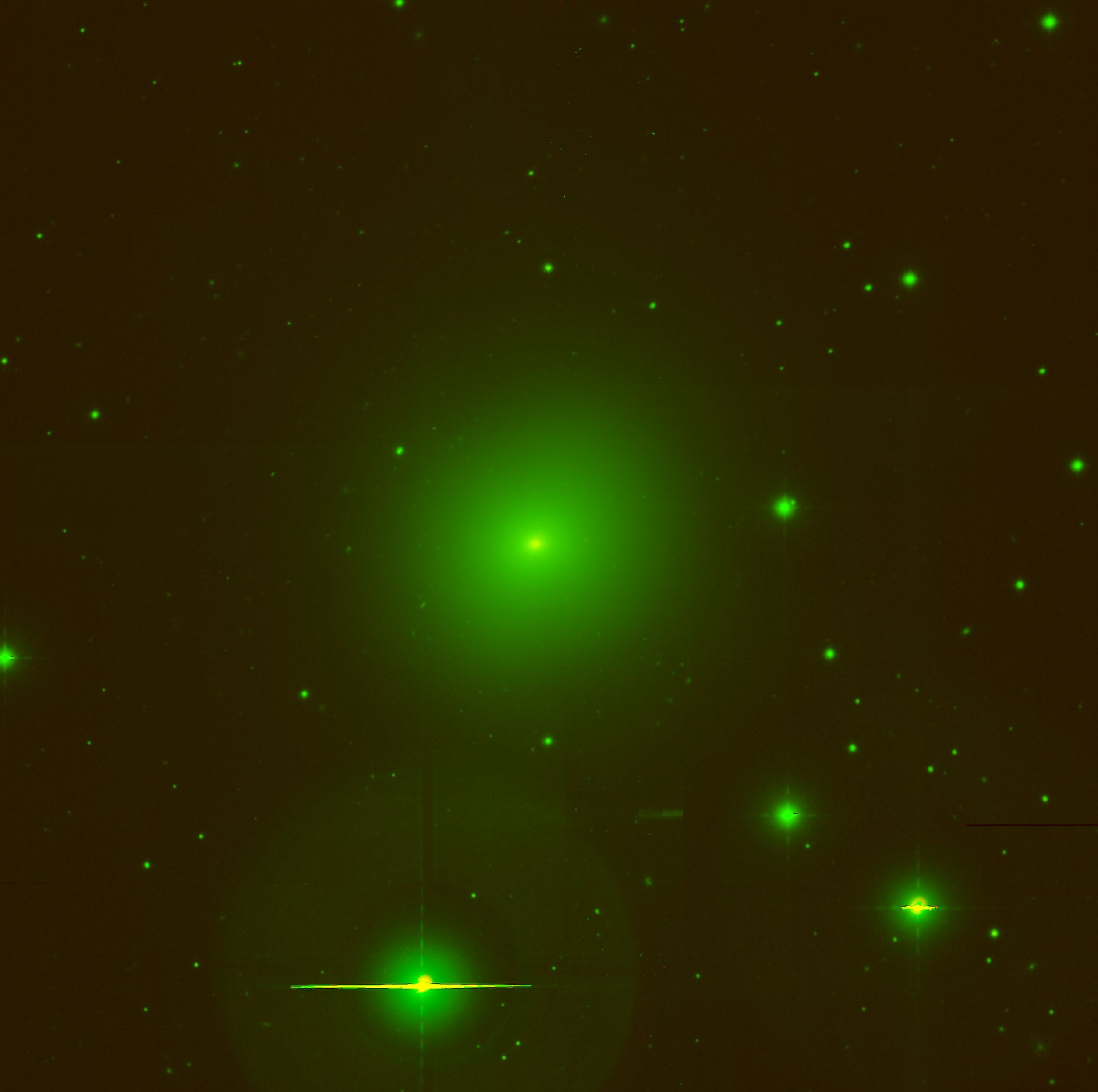}}
\caption{As in Figure~\ref{IC2038} for  NGC 1549
The image size is  8\arcmin$\times$8\arcmin.} 
\label{NGC1549}
\end{figure*}
%
\begin{figure*}[t]
\center
{\includegraphics[width=15cm]{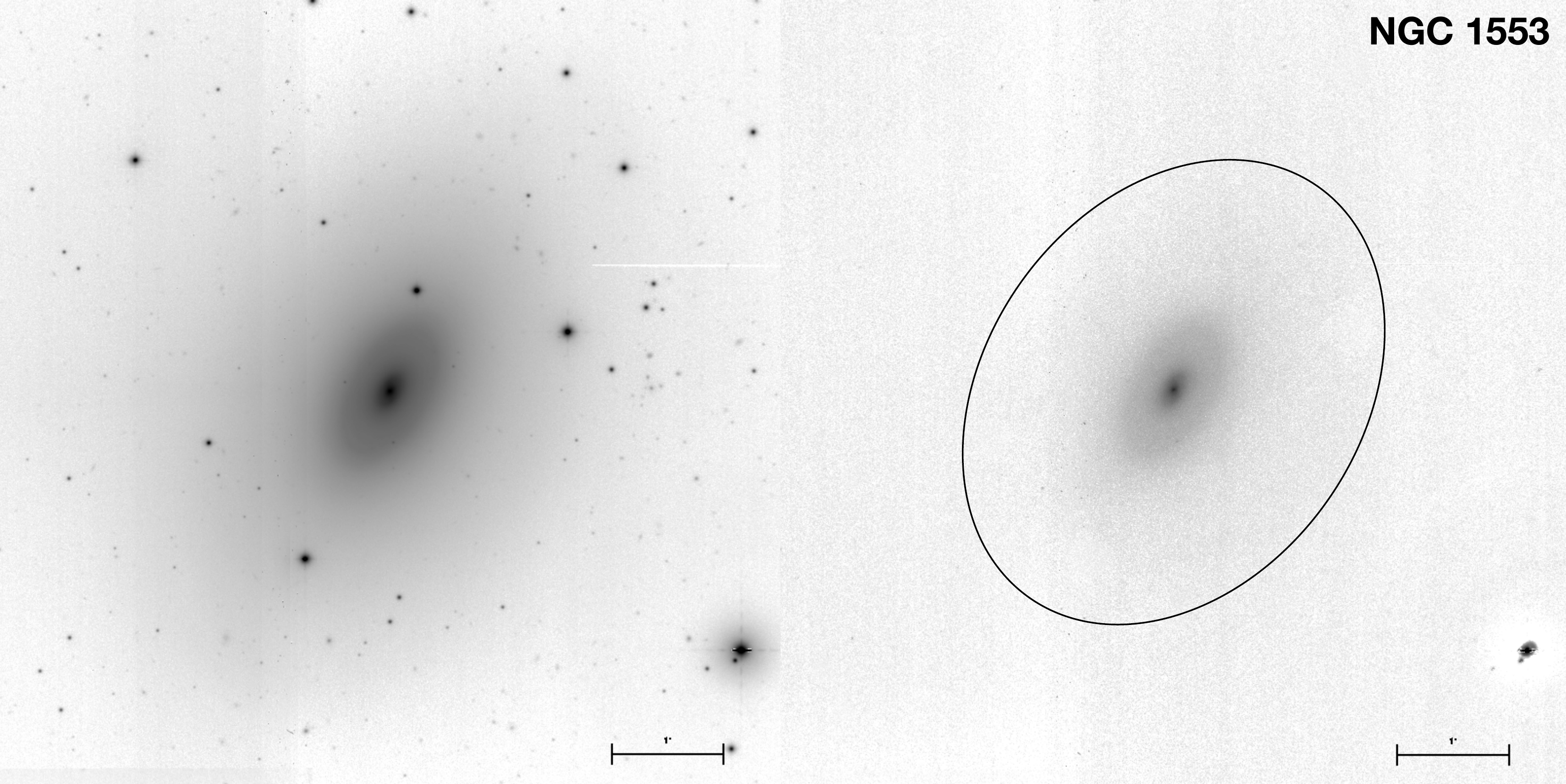}}
{\includegraphics[width=15cm]{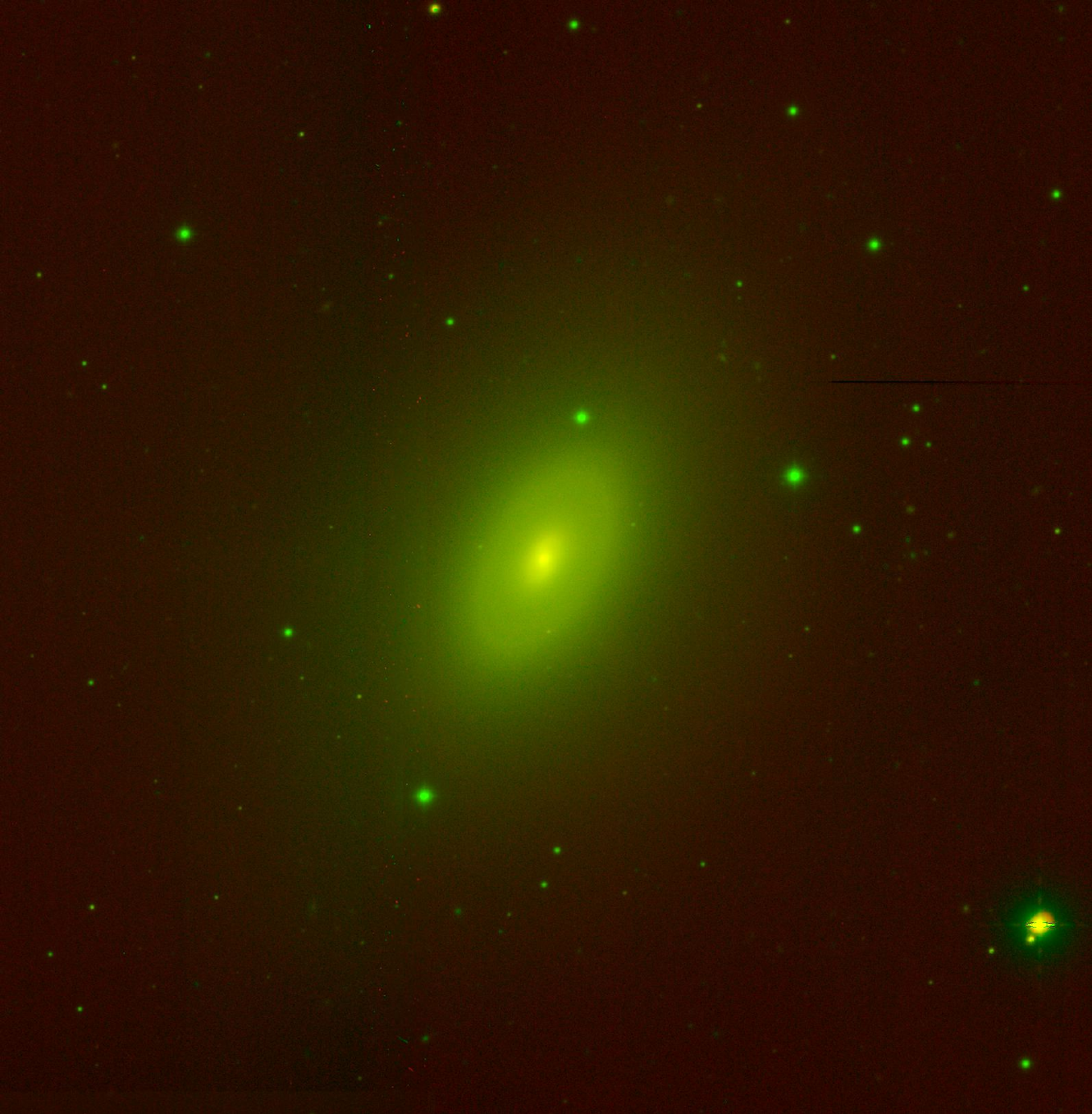}}
\caption{As in Figure~\ref{IC2038} for NGC 1553.
The image size is  7\arcmin$\times$7\arcmin. 
Two \Ha-bright regions are shown in detail in Figure~\ref{CMI2001}.} 
\label{NGC1553}
\end{figure*}
%
%

\begin{figure*}[t]
	\center
{\includegraphics[width=15cm]{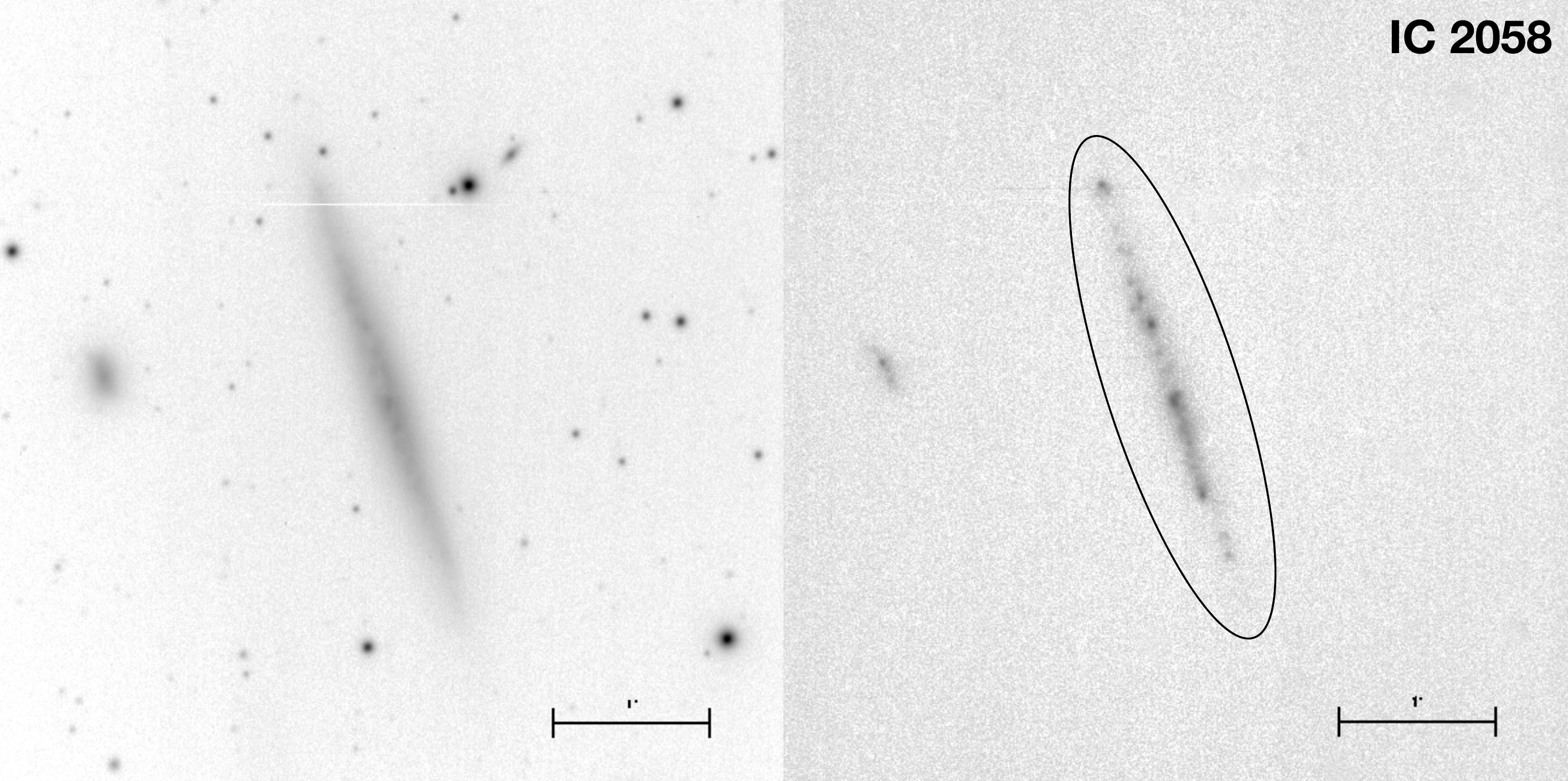}}
{\includegraphics[width=15cm]{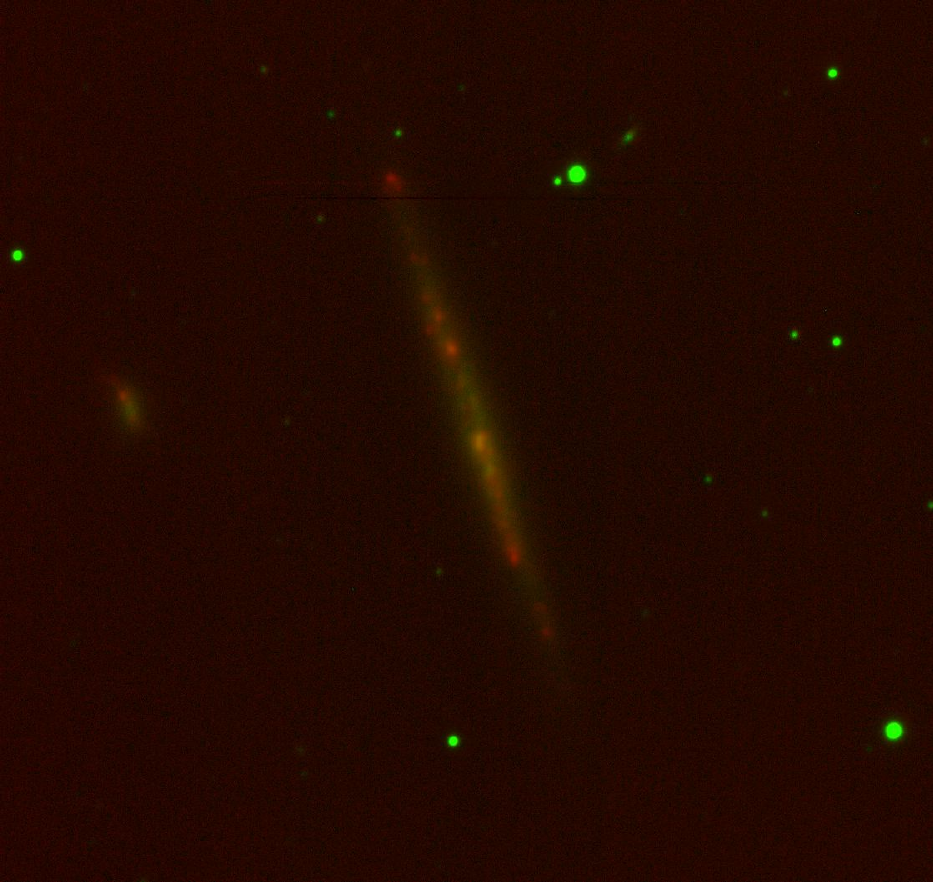}}
\caption{As in Figure~\ref{IC2038} for IC 2058
The image size is 5\arcmin$\times$5\arcmin. 
The physical companion of IC 2058, PGC 75125, 
is visible East of the galaxy.}
		\label{IC2058}
\end{figure*}

\begin{figure*}[t]
	\center	
	{\includegraphics[width=15cm]{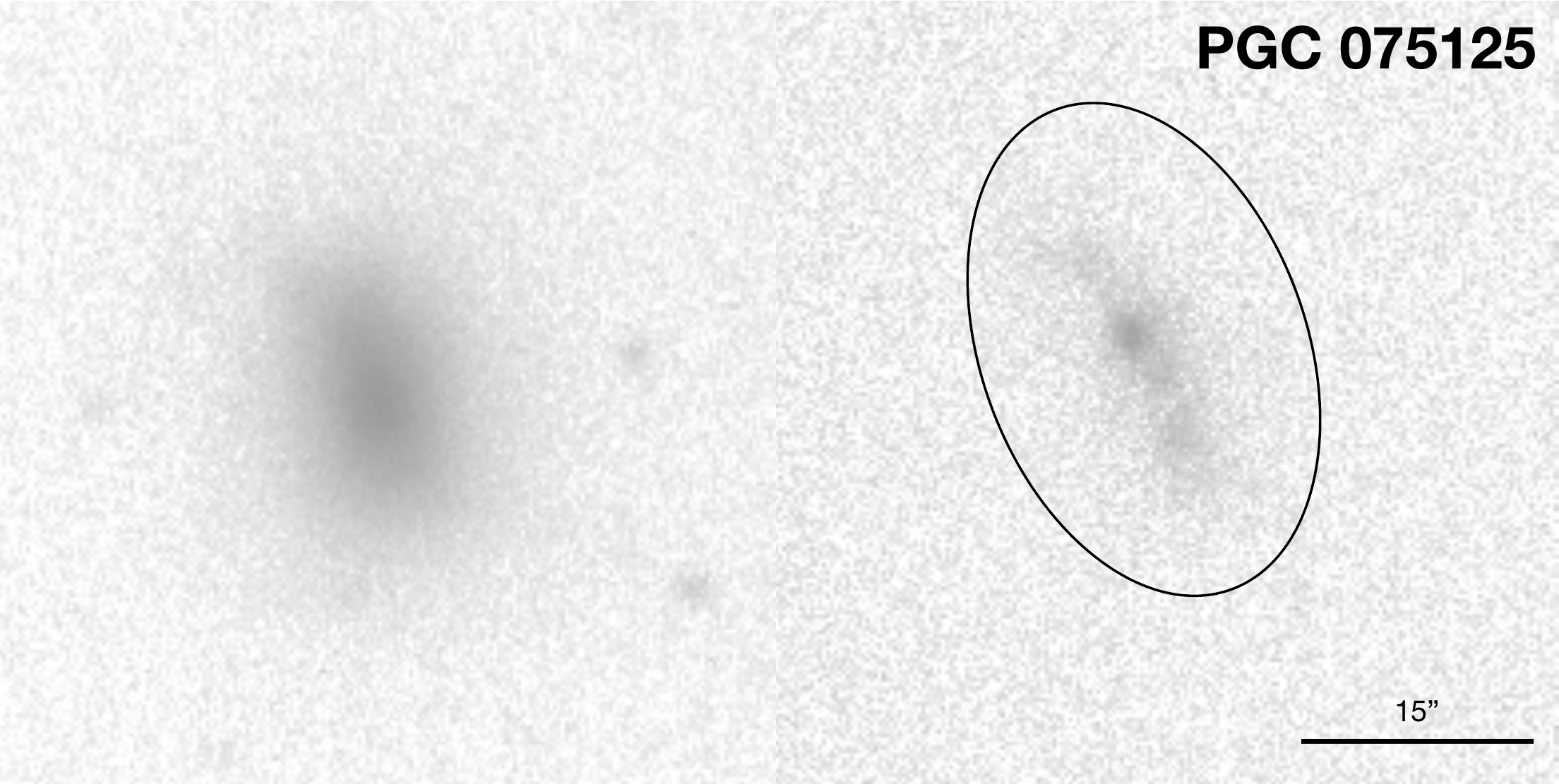}}
		{\includegraphics[width=15cm]{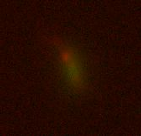}}
	\caption{As in Figure~\ref{IC2038} for PGC~ 75125. The image size is 
		50\arcsec$\times$50\arcsec. 
		The galaxy is classified as spiral (T=5$\pm3$) in {\tt HyperLeda}. In the
		continuum the galaxy structure appears boxy and without spiral arms. 
		The H$\alpha$+[NII] image shows a warped, clumpy star forming lane 
		crossing the galaxy. 
		\label{PGC075125}}
\end{figure*}

%
\begin{figure*}[t]
\center
{\includegraphics[width=15cm]{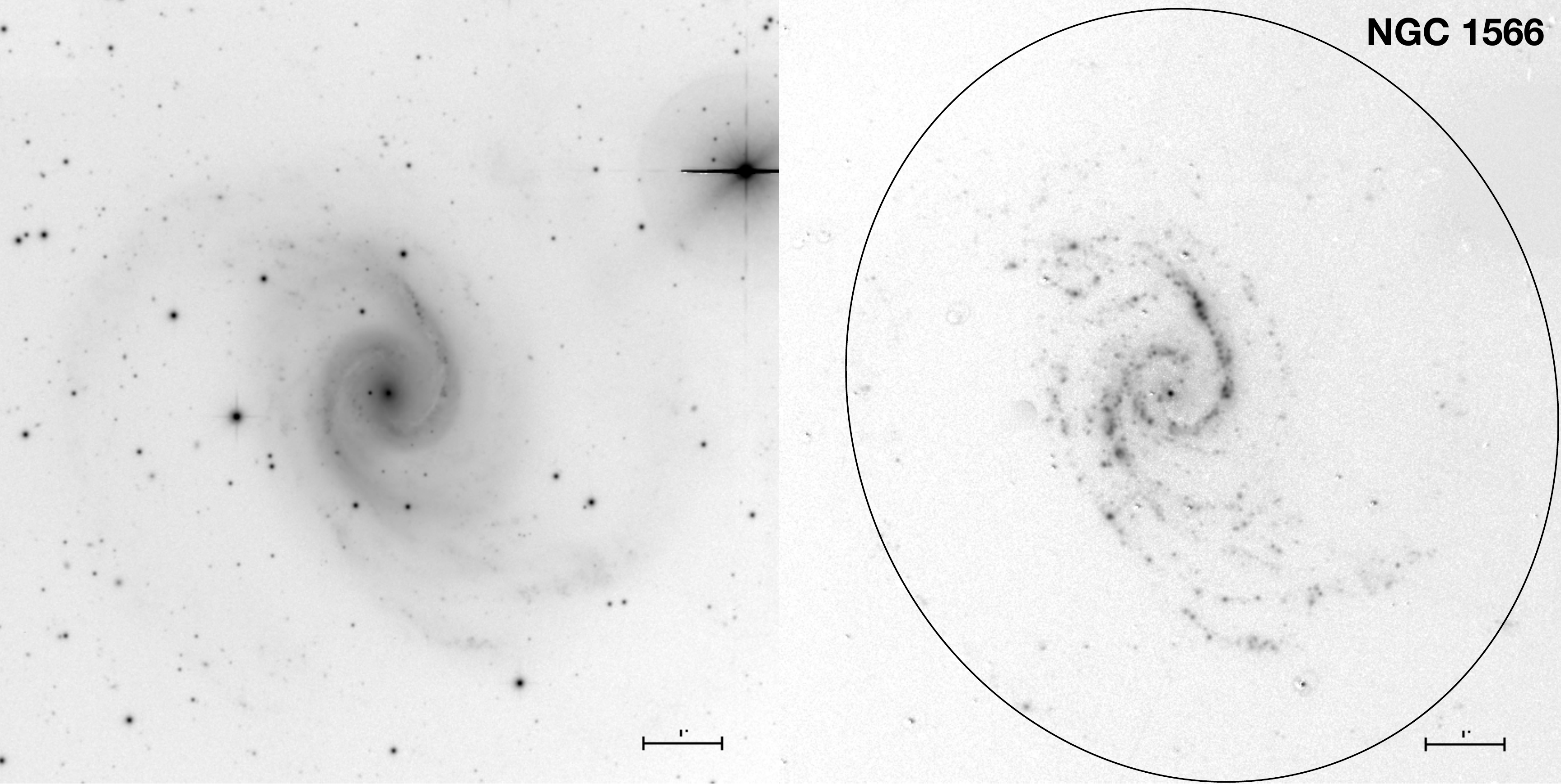}}
{\includegraphics[width=15cm]{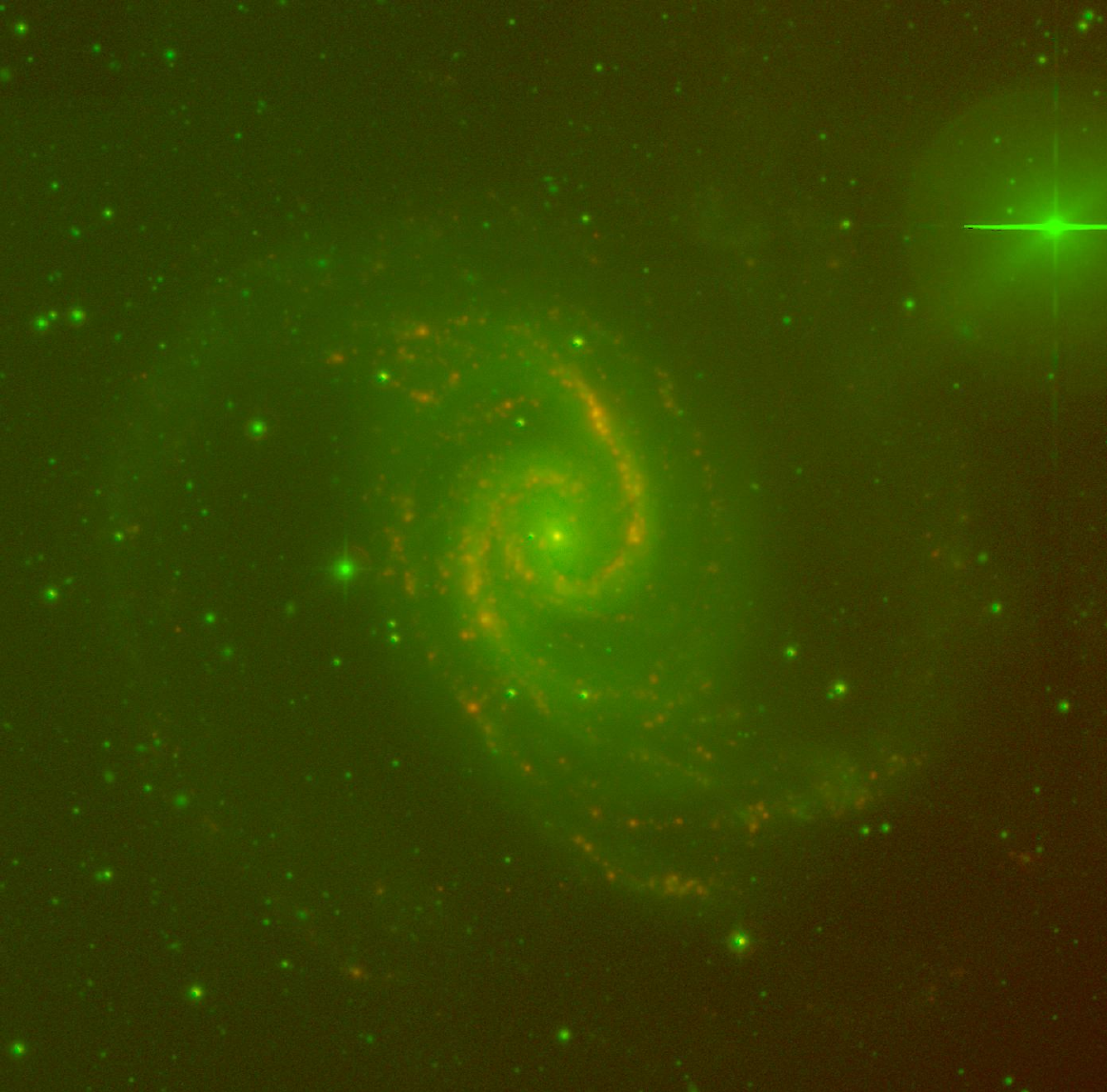}}
\caption{As in Figure~\ref{IC2038} for NGC 1566.
The image size is  10\arcmin$\times$10\arcmin.} 
\label{NGC1566}
\end{figure*}
%
\begin{figure*}[t]
\center
{\includegraphics[width=15cm]{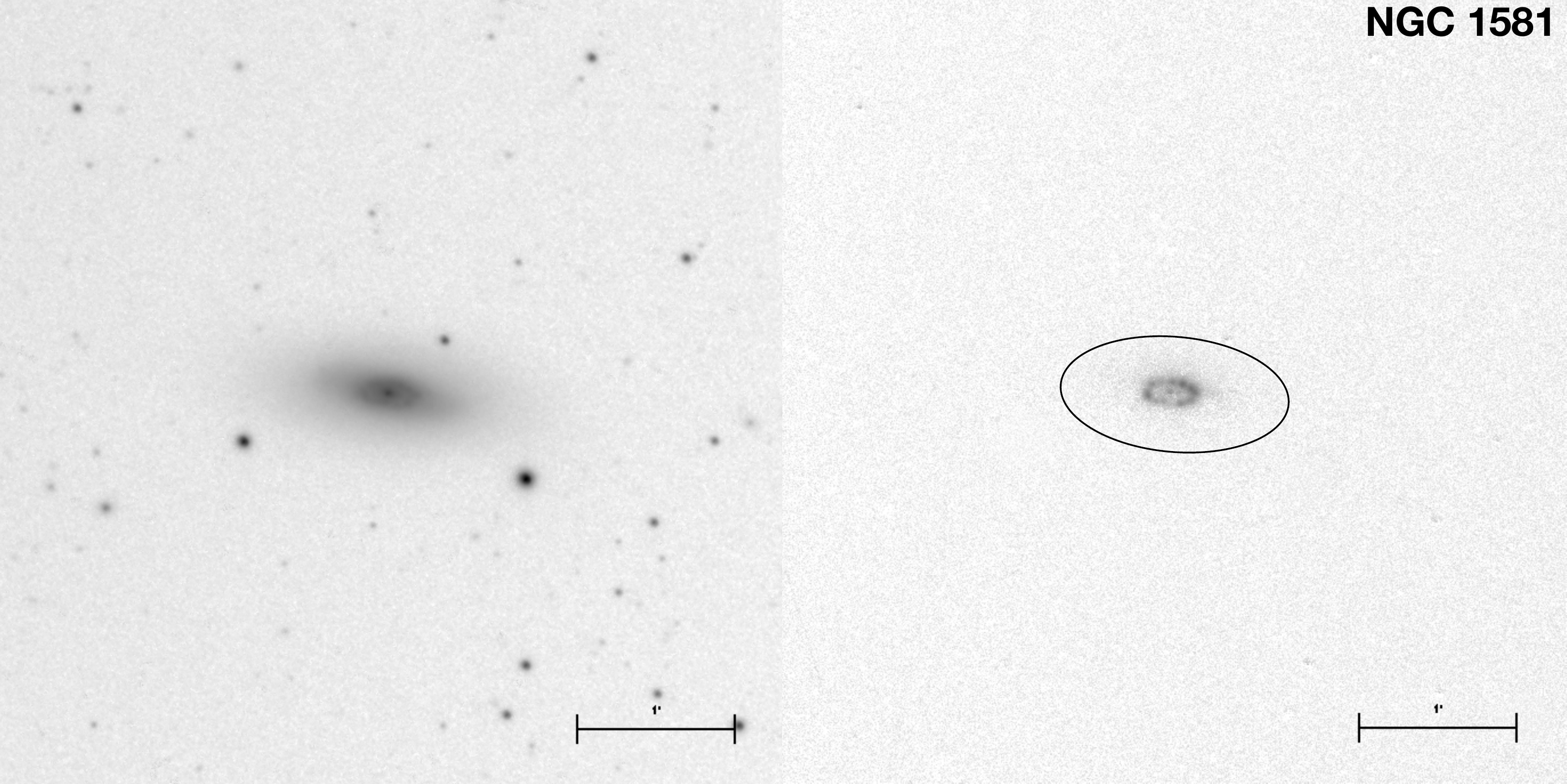}}
{\includegraphics[width=15cm]{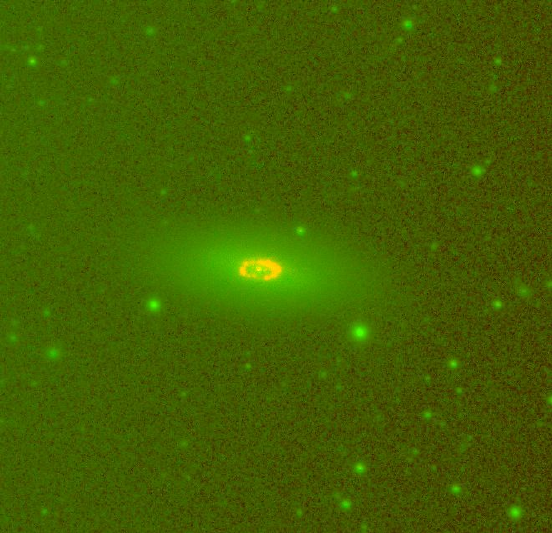}}
\caption{As in Figure~\ref{IC2038} for NGC 1581.
The image size is  5\arcmin$\times$5\arcmin.} 
\label{NGC1581}
\end{figure*}
%
%
\begin{figure*}[t]
	\center
{\includegraphics[width=15cm]{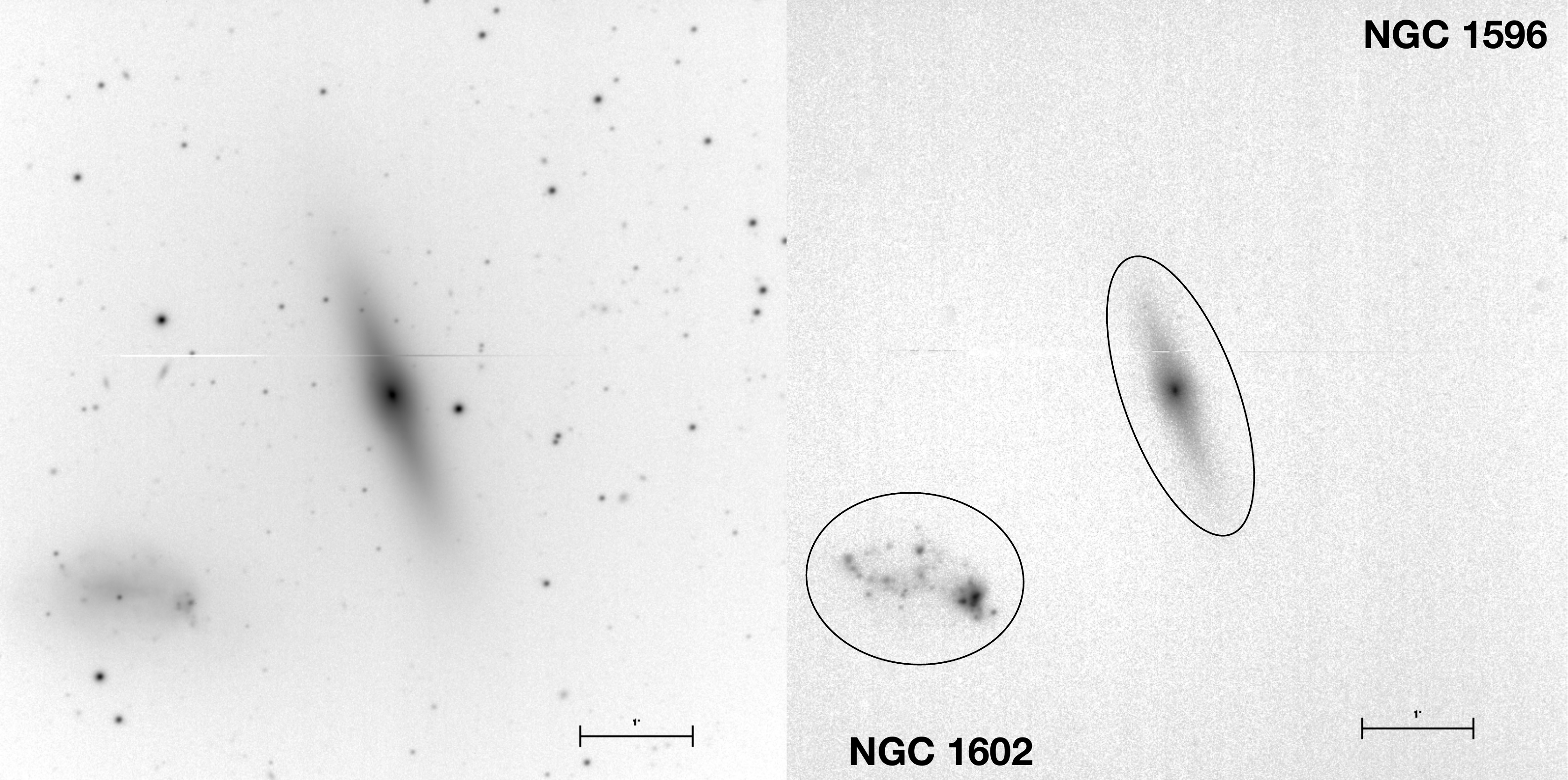}}
{\includegraphics[width=15cm]{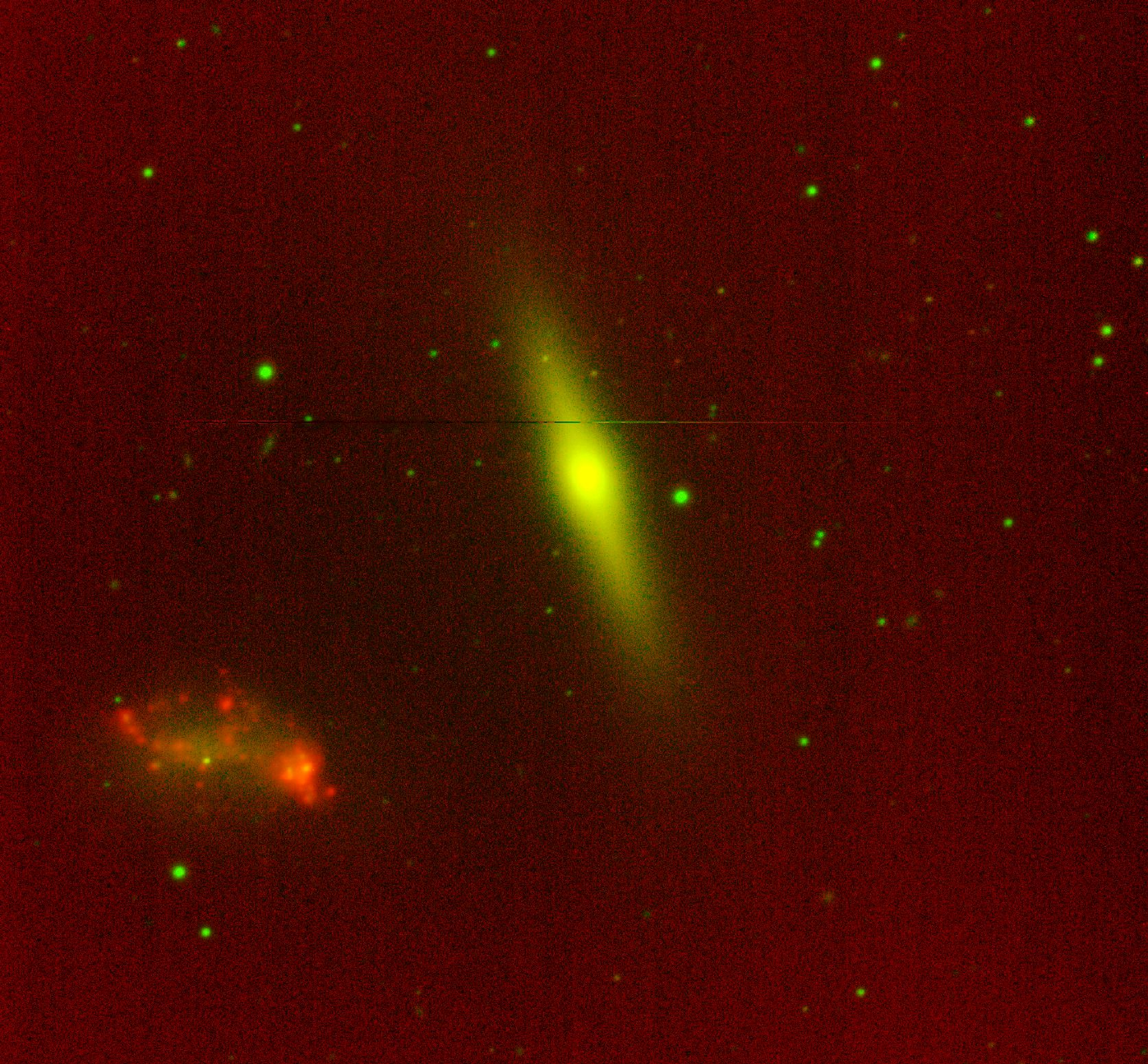}}
\caption{As in Figure~\ref{IC2038} for the physical pair \citep{Bureau2006} formed by 
NGC 1596 (NW) and NGC1602 (SE). The image size is  7\arcmin$\times$7\arcmin.}
\label{NGC1596}
\end{figure*}

%

\end{appendix}

\end{document}